\documentclass[oneside,12pt]{amsart}
\usepackage[T1]{fontenc}
\usepackage[utf8]{inputenc}
\usepackage{verbatim}
\usepackage{float}
\usepackage{amstext}
\usepackage{amsthm}
\usepackage{amssymb}
\usepackage{stmaryrd}
\usepackage{graphicx}

\makeatletter
\numberwithin{equation}{section}
\numberwithin{figure}{section}

\usepackage{amsfonts}
\usepackage[font=small,labelfont=bf]{caption}
\usepackage[unicode=true,
 bookmarks=true,bookmarksnumbered=false,bookmarksopen=false,
 breaklinks=false,pdfborder={0 0 1},backref=false,colorlinks=false]
 {hyperref}

\makeatother

\usepackage{babel}
\begin{document}
\title{QED positronium and the exact renormalization group}
\author{Guillermo Hansen$^{1}$ and Roberto Trinchero$^{2}$}
\date{25/06/2025}
\address{Instituto Balseiro and Centro Atómico Bariloche, 8400 Bariloche, Argentina}
\thanks{$^{1}$Supported by CONICET, $^{2}$Supported by CONICET and ANSES}
\begin{abstract}
Quantum electrodynamics (QED) is studied in the framework of the exact
(functional) renormalization group (ERG). This is done using an approach
to these equations which employs dimensional regularization. Simultaneous
solutions of the ERG equations and Ward identities are considered.
An approach to the study of bound states at long distances in the
ERG is employed to describe the positronium. This is done by introducing
a real scalar field which describes that bound state. The flow equations
are studied using an effective action ansatz and a low-momentum expansion.
\end{abstract}

\maketitle

\section{Introduction\label{sec:Introduction}}

The renormalization group ideas employed in field theory and statistical
mechanics constitute a breakthrough\cite{Wilson:1973jj} in terms
of the understanding of the general structure of the space of quantum
field theories. In addition the so called exact or functional renormalization
group (ERG) equations give a very precise formulation of that ideas\cite{Weinberg1978}\cite{WETTERICH199390}.
From the axiomatic point of view, in Euclidean space it can be said
that a field theory exists if it is possible to find a solution of
the ERG equations which satisfies the Osterwalder-Schräder axioms\cite{Osterwalder1973}.

The formulation of gauge theories in this scheme is quite involved.
This is so because the original formulation employs cut-off procedures
which do not respect gauge symmetries. It is remarkable that the same
ideas can be formulated in terms of dimensional regularization which
respects gauge symmetries\cite{Trinchero_2022}.

In the present work the above mentioned approach is applied to QED.
The main motivation is to test these ideas in a concrete physical
system. This is done taking into account the existence of the positronium
bound state in this theory\footnote{The description of bound states within the ERG approach has been considered
by many authors \cite{Ellwanger_1994}\cite{Gies_2002}\cite{Floerchinger_2009}\cite{Alk19PhysRevD.99.054029}. }. An approach to the long distance description of bound states within
the ERG framework is given and applied to the description of the positronium
in QED. 

The features and results of this work are summarized as follows,
\begin{itemize}
\item Fields are included for both point particles and bound states. The
positronium at long distances is described by a real scalar field
$\phi$.
\item The positronium scalar field is related to the electron-positron field
$\psi$ by,
\[
\tilde{m}\phi=\frac{1}{g}\bar{\psi}\psi
\]
this equation is implemented by means of a delta function. This entails
the appearance of a mass term for the positronium scalar field, a
Yukawa coupling of this field with the $\psi$ field and a four fermion
interaction.
\item The ERG for the corresponding theory are considered. These equations
are supplemented by the Ward-Takahashi (WT) identities imposing gauge
symmetry at the level of the proper functions. This means that simultaneous
solutions to both the ERG equations and the WT identities are considered.
\item An approximation of the resulting equations is implemented. This approximation
consists of a truncation, which leaves only a finite number of proper
functions, and a low-momentum expansion. In order to describe the
dynamics of the positronium, the surviving proper functions after
truncation include a kinetic term and a potential for the positronium
scalar field. This potential only keeps engineering-relevant monomials
in the field $\phi$, that is, with momentum dimension less than or
equal to $4$.
\item The ERG equations for the surviving proper functions after truncation
are obtained in this approximation. This involves the calculation
of a number of Feynman graphs. The internal lines in these graphs
are given by complete propagators for the gauge field, the electron-positron
field, and the positronium scalar field. The vertices are given by
the proper functions. 
\item Anomalous dimensions are computed in this approximation by requiring
unit coefficients for the kinetic terms of all the fields.
\item The electron mass is generated by a non-vanishing expectation value
for the positronium field via the Yukawa term. 
\item The positronium corresponds to the excitation of the scalar field
around the non-trivial vacuum of this field.
\item The positronium decay is computed using the first diagram contributing
to its decay into two photons.
\item The initial conditions for the ERG equations are obtained by fixing
the electron mass, the electron charge, the positronium mass, and
its decay exponent. 
\item The ERG equation are numerically solved using the initial conditions
mentioned above. This fixes the value of all the considered proper
functions for long distance scales. This amounts to having a long-distance
effective theory describing electron, positrons, and positronium interactions
at long distances.
\item The coincidence of the results with the running of the mass and charge
of the electron in QED at long distances without the scalar field
holds up to energies of the order of $150$ times the value of the
electron rest mass. 
\end{itemize}
This paper is organized as follows. Section 2 presents the effective
action to be considered and implements the delta function mentioned
in the introduction. In Section 3, the ERG equations and the calculation
of the anomalous dimensions are described. Section 4 deals with the
approximations employed. Section 5 shows how the electron mass is
generated, describes the positronium mass and decay exponent, the
initial conditions for the ERG equations, and the solutions of the
ERG equations. Finally, Section 6 presents some conclusions and outlook.
The main text is supplemented by seven appendices. These appendices
include a derivation of the ERG equations\footnote{This derivation shows the equivalence between the approach in ref.
\cite{Weinberg1978} and the one in \cite{WETTERICH199390}.}(Appendix \ref{Derivation}) and the explicit example of the $\phi^{4}$
theory in this approach (Appendix \ref{fi4}). In addition several
technical points not included in the main text are dealt with in the
other appendices.

\section{The effective action including bound states\label{sec:The-effective-action}}

In the ERG approach to a field theory which may present bound states,
there are various objects to be considered. The fields representing
the point particles and bound states should be chosen and also the
effective action terms to be studied. However, it is clear that the
notions of bound state and point particle depend on the scales the
observer can resolve. Indeed, at large distances from the observer
the bound state behaves as a point particle. Then it seems useful
to include fields for all the particles composite or not. However,
fields corresponding to point particles and fields corresponding to
composite particles should be related. For the case of a composite
real scalar with two fermion constituents, the corresponding relation
would involve two fermion fields and, in order to preserve gauge invariance,
a Wilson line joining them. This is a non-trivial object whose description
is quite involved. However the description of this composite object
at long distances is much simpler. At long distances, the composite
object behaves as a point particle and the two fermions can be considered
to be at the same point. This effective long distance description
can be implemented using the identity
\begin{align}
\prod_{x}\,\delta(\tilde{m}\phi-\frac{1}{g}\bar{\psi}\psi) & =\lim_{\lambda\to0}\frac{1}{\sqrt{\lambda}}\,e^{-\frac{1}{2\lambda}\int d^{d}x\,\left(\tilde{m}\phi-\frac{1}{g}\bar{\psi}\psi\right)^{2}}\label{eq:deltafipsi}
\end{align}
The parameters $\tilde{m}$ and $g$ are dimension-one constants,
and $\lambda$ is dimensionless. The above identity include a mass
term for the field $\phi$ with mass $m_{\phi}=\tilde{m}/\sqrt{\lambda}$,
a four-fermion interaction term with coupling constant $\lambda_{4f}=1/2\lambda g^{2}$
and cross Yukawa terms, with dimensionless coupling $\lambda_{Y}=-\tilde{m}/\lambda g$.
It is noted that these couplings are not independent; they are related
by,
\begin{equation}
\lambda{}_{4f}=\frac{\lambda_{Y}^{2}}{2m_{\phi}^{2}}\,.\label{eq:delta-coup}
\end{equation}
It is emphasized that this relation, as the whole approach, is expected
to be valid only at long distances. These coupling constants will
have a dependence on the scale, which is determined by the ERG equations.
The propagation and interactions of the composite particles among
themselves can be described by including a kinetic term $\frac{1}{2}\,\partial_{\mu}\phi\,\partial_{\mu}\phi^{\dagger}$
and a potential of the form $V(\phi^{\dagger}\phi)$. Only the relevant
$\phi^{3}$ and $\phi^{4}$ terms in this potential will be considered.
All this leads to the following ansatz for the effective action\footnote{The conventions employed for the euclidean gamma matrices, the Dirac
action and the Dirac propagator are described in Appendix A \ref{sec:Appendix-A.-Gamma}.}:
\begin{align}
\Gamma[A_{\mu},\psi,\bar{\psi},\phi] & =\int d^{4}x\,\Big(-\frac{1}{4}F_{\mu\nu}F_{\mu\nu}+\bar{\psi}\,\Big(\gamma^{\mu}(\partial_{\mu}-ieA_{\mu})-m\Big)\,\psi(x)\nonumber \\
 & +\frac{1}{2}\partial_{\mu}\phi\partial_{\mu}\phi+\frac{\lambda_{4}}{4!}\phi^{4}+\frac{\lambda_{3}}{3!}\phi^{3}\Big)+\frac{1}{2\lambda}\int d^{d}x\,\left(\tilde{m}\phi-\frac{1}{g}\bar{\psi}\psi\right)^{2}\,.\label{eq:efac}
\end{align}
The masses and coupling constants appearing in the above effective
action are,
\[
m,\,\tilde{m},\,e,\,\lambda_{3},\,\lambda_{4},\text{\,}\lambda_{Y},\,\lambda_{4f}\,.
\]
The general form of this effective action is
\begin{equation}
\Gamma[A,\bar{\psi},\psi,\phi]=\int\frac{1}{n!m!l!}\,\sum_{n,m,l}\Gamma^{n,2m,l}\cdot\underset{n}{\underbrace{A\cdots A}}\:\underset{2m}{\underbrace{\bar{\psi}\psi\cdots\bar{\psi}\psi}}\:\underset{l}{\underbrace{\phi\cdots\phi}\,.}\label{eq:efac-gral}
\end{equation}
In the above expression, the dot $\cdot$ between the proper functions
and the fields indicates contraction of the corresponding indices;
the integral indicates integration over all the momentum variables.
The proper two-point functions at zero momentum correspond to the
masses appearing in the ansatz~(\ref{eq:efac}), that is,
\begin{align*}
m & =\Gamma^{0,2,0}(0)\\
\frac{\tilde{m}^{2}}{\lambda} & =\Gamma^{0,0,2}(0)
\end{align*}
The coefficients of the interaction terms appearing in~(\ref{eq:efac})
correspond to some of the proper functions appearing in~(\ref{eq:efac-gral})
at zero momentum, as given by the following identities,
\begin{align*}
\lambda_{4} & =\Gamma^{0,0,4}(0,0,0)\,,\\
\lambda_{3} & =\Gamma^{0,0,3}(0,0)\,,\\
e\gamma_{\mu} & =\Gamma_{\mu}^{1,2,0}(0,0)\,,\\
\frac{1}{2\lambda g^{2}} & =\Gamma^{0,4,0}(0,0,0)\,,\\
-\frac{\tilde{m}}{\lambda g} & =\Gamma^{0,2,1}(0,0)\,.
\end{align*}

\section{The flow equations}

In this section, the ERG equations for the terms appearing in the
effective action ansatz in~(\ref{eq:efac}) are considered\footnote{The derivation of this equation is considered in Appendix \ref{Derivation}.}.
The general form of these equations is
\[
\left[\mu\frac{\partial}{\partial\mu}-n\frac{\gamma_{A}}{2}-m\gamma_{\psi}-l\frac{\gamma_{\phi}}{2}\right]\,\Gamma^{n,2m,l}(p_{1},\cdots,p_{n+2m+l})=L_{}^{n,2m,l}(\{p\};\mu)
\]
and $\gamma_{A},\gamma_{\psi}$ and $\gamma_{\phi}$ indicate the
anomalous dimensions of the corresponding fields. The right-hand side~(r.h.s.)
of these equations are of the following general form:
\begin{align*}
L_{}^{n,2m,l}(\{p_{1},\cdots,p_{n+2m+l};\mu) & =-\frac{1}{2}\left[\partial_{t}\bar{G}_{>ij}\cdot\sum_{l=0}^{\infty}(-1)^{l}\sum_{m_{1}\cdots m_{l}}\frac{\delta^{m_{1}}}{\delta\phi(p_{1})\cdots\delta\phi(p_{m_{1}})}\left(\frac{\delta^{2}\bar{\Gamma}_{\Lambda}[\phi]}{\delta\phi\delta\phi}\right)_{jk_{1}}\right.\\
 & \left.\cdots\:\bar{G}_{>}\cdot\frac{\delta^{m_{l}}}{\delta\phi(p_{k-m_{l}})\cdots\delta\phi(p_{m_{l}})}\left(\frac{\delta^{2}\bar{\Gamma}_{\Lambda}[\phi]}{\delta\phi\delta\phi}\right)_{k_{l-1}i}\right]_{\phi=0}
\end{align*}
where a multi-index notation is employed. That is, the indices $i,j,k,\cdots$
indicate type of field, component of field and momentum. The indices
$m_{1}\cdots m_{l}$ are such that,
\[
m_{1}+m_{2}+\cdots+m_{l}=n+2m+l
\]
Diagrammatically, the $L^{n,2m,l}(\{p\};\mu)$ are represented by
one-loop diagrams.

\begin{figure}
\vspace{3cm}

\begin{raggedright}
\includegraphics[bb=0bp 25bp 330bp 270bp]{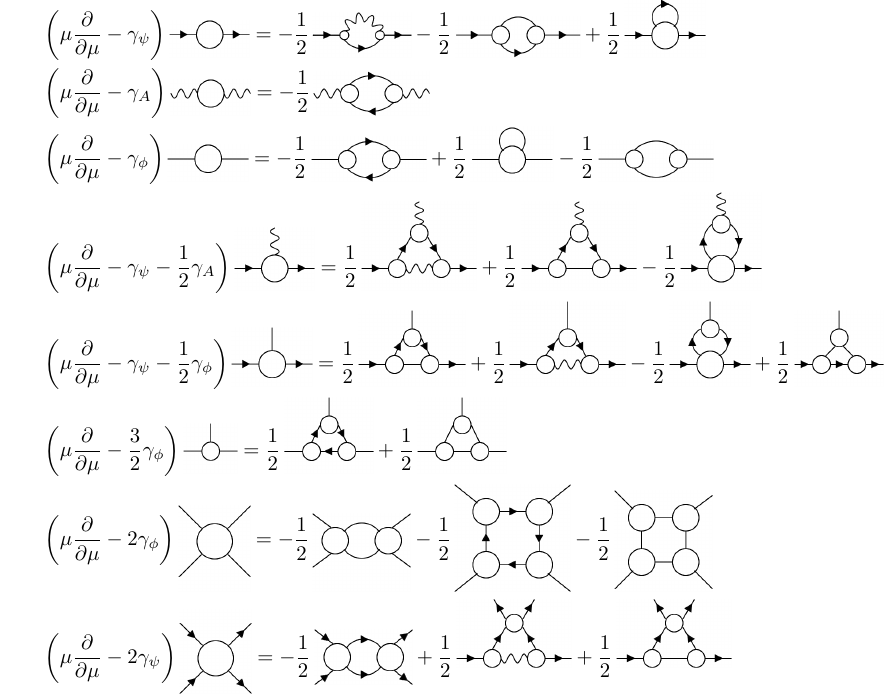}
\par\end{raggedright}
\vspace{1cm}

\caption{The ERG equations for the proper functions appearing in (\ref{eq:efac}).\label{fig:erg}}
\label{figerg}
\end{figure}

These equations can be represented as in Figure~\ref{fig:erg} .

The internal lines in these graphs correspond to photon (wavy lines),
fermion (arrowed lines) and scalar (solid lines) propagation\footnote{The general expressions used in the $1$-loop integrals appearing
in the above graphs, which involve the $\theta_{\epsilon}$ function,
are given in Appendix~\ref{sec:Appendix-B.-Integrals} according
to the results in ref. \cite{Trinchero_2022}.}. They represent the following complete propagators:
\begin{align}
G(k) & =\theta_{\epsilon}(\frac{k^{2}}{\mu^{2}}-1)\,\frac{g_{\mu\nu}}{k^{2}+\Sigma_{P}(k^{2})},\quad S(p)=\theta_{\epsilon}(\frac{p^{2}}{\mu^{2}}-1)\,\frac{1}{i\gamma^{\mu}p_{\mu}-m+\Sigma_{F}(p)}\label{eq:prop-1}\\
D(q) & =\theta_{\epsilon}(\frac{q^{2}}{\mu^{2}}-1)\,\frac{1}{q^{2}+\frac{\tilde{m}^{2}}{\lambda}+\Sigma_{\phi}(q^{2})}=\theta_{\epsilon}(\frac{q^{2}}{\mu^{2}}-1)\,\frac{\lambda}{\lambda q^{2}+\tilde{m}^{2}+\lambda\Sigma_{\phi}(q^{2})}
\end{align}
Using the complete propagators has as a consequence that no two-point
proper vertices appear in the diagrams~\cite{Weinberg1978}. 

\noindent The proper functions are represented by empty blobs in the
above graphs. They will be denoted by,
\[
\Gamma^{n,m,l}\,,
\]
where $n,\,m,\,l$ respectively denote the number of photon legs,
fermion legs and real scalar legs\footnote{The dimensions of these proper functions are,
\[
[\Gamma^{n,m,l}]=d-n\,\left(\frac{d-2}{2}\right)-m\,\left(\frac{d-1}{2}\right)-l\,\left(\frac{d-2}{2}\right)\,,
\]
which gives,
\begin{align*}
[\Gamma^{2,0,0}] & =2=[\Gamma^{0,0,2}]\;,\;[\Gamma^{0,2,0}]=1\,,\\{}
[\Gamma^{1,2,0}] & =\frac{\epsilon}{2}=[\Gamma^{0,2,1}]\,,\\{}
[\Gamma^{0,0,4}] & =\epsilon\;,\;[\Gamma^{0,4,0}]=\epsilon-2\,,
\end{align*}
this implies that,
\[
[e]=\frac{\epsilon}{2}=\left[\frac{2\tilde{m}}{\lambda g}\right]\;,\;\left[\frac{1}{\lambda g^{2}}\right]=\epsilon-2\,.
\]
}

\subsection{Gauge invariance }

As mentioned in Section \ref{sec:Introduction}, solutions to both
the ERG equations and the WT identities are considered. The ERG equations
that are non-trivially affected by the WT identities are the ones
involved in the first, second and forth equations appearing in fig.
\ref{fig:erg}. The second equation involves the vacuum polarization
tensor $\Gamma_{\mu\nu}^{2,0,0}(k)$, which should be transversal
as the following WT shows, i.e.,
\begin{equation}
k^{\mu}\Gamma_{\mu\nu}^{2,0,0}(k)=0\,.\label{eq:tr}
\end{equation}
To see that the second ERG equation in Fig. \ref{fig:erg} is consistent
with the WT identity above, both sides of the ERG equation are contracted
with $k^{\mu}$, the left-hand side (l.h.s.) vanishes because of (\ref{eq:tr}).
Regarding the r.h.s., employing the WT identity which involves the
QED vertex correction,
\begin{equation}
(p-q)^{\mu}\Gamma_{\mu}^{1,2,0}(p,q)=\Gamma^{0,2,0}(p)-\Gamma^{0,2,0}(q)\,,\label{eq:wtvertex}
\end{equation}
shows that it is consistent with l.h.s., provided that it is regularized
with a gauge invariant regulator. This is the case in this approach,
since it is regularized using dimensional regularization. It is remarked
that the transversal form of $\Gamma_{\mu\nu}^{2,0,0}$, implied by
the WT identity (\ref{eq:tr}), will be employed in the calculation
of the anomalous dimension $\gamma_{A}$.

The WT identity (\ref{eq:wtvertex}) also relates the first and fourth
equations in Fig. \ref{fig:erg}. Indeed, contracting the l.h.s of
the fourth equation with $(p-q)^{\mu}$ leads to the l.h.s. of the
first equation, and contracting the r.h.s of the fourth equation with
$(p-q)^{\mu}$ leads to the r.h.s of the first equation. Thus, if
the first equation holds, then the fourth is consistent with the WT
identity (\ref{eq:wtvertex}). It is also worth noting that, due to
the WT identity (\ref{eq:wtvertex}), there is a cancellation between
the first term on the r.h.s. of the fourth equation and the anomalous
dimension $\gamma_{\psi}$.

\subsection{Anomalous dimensions}

It is important to state that fields will be re-scaled so that the
coupling $K_{A}$ , $K_{\psi}$ and $K_{\phi}$ corresponding to the
kinetic terms $Q_{A}$, $Q_{\psi}$ and $Q_{\phi}$ are equal to $1$
for any scale\footnote{It is worth noting that this condition implies that the expansion
of the proper two point functions in momentum up to the first non-trivial
contribution, is taken into account in the calculation. } $\mu$, i.e.,
\begin{align*}
Q_{A} & =-\frac{1}{2}\,K_{A}\,A^{\mu}(k)\,(g_{\mu\nu}k^{2}-k_{\mu}k_{\nu})\,A^{\nu}(-k)\,,\quad K_{A}=1\,,\\
Q_{\psi} & =K_{\psi}\,(i\gamma^{\mu}p_{\mu}-m)\quad,\quad K=1\,,\\
Q_{\phi} & =\frac{1}{2}\,K_{\phi}(q^{2}+m^{2})\quad,\quad K_{\phi}=1\,.
\end{align*}
This entails a redefinition of the proper functions so that the effective
action (\ref{eq:efac}) remains the same. The above conditions in
terms of the two-point proper functions are,
\begin{align}
\left.\frac{\partial}{\partial p_{1}^{\nu}}\Gamma^{0,2,0}(p_{1};\mu)\right|_{p_{1}=0} & =i\gamma_{\nu}\,,\nonumber \\
\left.\frac{\partial^{2}}{\partial k^{\rho}\partial k^{\sigma}}\Gamma_{\mu\nu}^{2,0,0}(k;\mu)\right|_{k=0} & =g_{\rho\sigma}g_{\mu\nu}-\frac{1}{2}g_{\mu\rho}g_{\nu\sigma}-\frac{1}{2}g_{\mu\sigma}g_{\nu\rho}\,,\nonumber \\
\left.\frac{\partial^{2}}{\partial q^{\rho}\partial q^{\sigma}}\Gamma^{0,0,2}(q;\mu)\right|_{q=0} & =\frac{1}{2}g_{\rho\sigma}\,.\label{eq:kinetic-cond-1}
\end{align}
Taking the first derivative respect to $p$ of the first equation
in Fig.$(\ref{fig:erg})$, the second derivative respect to $q^{\rho}$
and $q^{\sigma}$ of the third equation in that figure, and applying
$g^{\rho\sigma}g^{\mu\nu}$ and the second derivative respect to $k^{\rho}$
and $k^{\sigma}$ of the second equation in that figure, and evaluating
and at $p=k=q=0$ and using (\ref{eq:kinetic-cond-1}) leads to,
\begin{align}
\gamma_{\psi} & =\frac{i}{16}\epsilon\int\frac{d^{d}p}{(2\pi)^{n}}\frac{\partial}{\partial q^{\eta}}\text{{Tr}}\left[\gamma^{\eta}\Gamma_{\nu}^{1,2,0}(q-p,p,-q)S(p)\Gamma_{\mu}^{1,2,0}(p-q,q,-p)G^{\mu\nu}(p-q)\right.\nonumber \\
 & \left.+\gamma^{\eta}\Gamma^{0,2,1}(q-p,p,-q)S(p)\Gamma^{0,2,1}(p-q,q,-p)D(p-q)\right]_{q=0}\,,\label{eq:gampsi}
\end{align}
\begin{align}
\gamma_{\phi} & =-\frac{1}{2}g_{\rho\sigma}\frac{\partial^{2}}{\partial q_{\rho}\partial q_{\sigma}}\left[\epsilon\int\frac{d^{d}p}{(2\pi)^{n}}\Gamma^{0,2,1}(q,p)S(p)\Gamma^{0,2,1}(q,q-p)S(q-p)\right.\nonumber \\
 & \left.+\,\epsilon\int\frac{d^{d}p}{(2\pi)^{n}}\Gamma^{0,0,3}(q,p)D(p)\Gamma^{0,0,3}(q,q-p)D(q-p)\right]_{q=0}\,,\label{eq:gammaphi}
\end{align}
\begin{align}
\gamma_{A} & =\frac{1}{12}g^{\mu\nu}\left(\frac{\partial}{\partial q}\cdot\frac{\partial}{\partial q}\right)\text{{Tr}}[\mu^{-\epsilon}\int\frac{d^{n}p}{(2\pi)^{n}}\Gamma_{\mu}^{1,2,0}(q,p,-(p+q))\left(\mu\frac{\partial}{\partial\mu}S(p)\right)\nonumber \\
 & \times\Gamma_{\nu}^{1,2,0}(-q,p+q,-p)S(p+q)]\,.\label{eq:gamma-A-1}
\end{align}

\begin{comment}
Due to gauge invariance the general form of $\Gamma_{\mu\nu}^{2,0,0}(k)$
is,
\[
\Gamma_{\mu\nu}^{2,0,0}(k)=(\delta_{\mu\nu}k^{2}-k_{\mu}k_{\nu})\,\Pi(k^{2})
\]
thus,
\[
\Gamma_{\mu\mu}^{2,0,0}(k)=(d-1)\,k^{2}\,\Pi(k^{2})
\]
 replacing in the first equation in Fig. \ref{fig:erg} leads to,
\[
(\partial_{t}-\gamma_{A})\,\Gamma_{\mu\mu}^{2,0,0}(k)=+\frac{1}{2}e^{2}\epsilon\int\frac{d^{d}p}{(2\pi)^{d}}\text{\,{Tr}}\left[\gamma_{\mu}\frac{(i\not p+m)}{p^{2}+m^{2}}\gamma_{\mu}\frac{(i\not p+m)}{p^{2}+m^{2}}\right]\,.
\]
\end{comment}

\section{Non-perturbative approximation\label{sec:Non-perturbative-approximation}}

The approximation consists first in restricting to the effective action
ansatz in (\ref{eq:efac}). That is, all proper functions corresponding
to interaction terms which do not appear in the effective action (\ref{eq:efac})
are neglected. Furthermore, a low momentum expansion for the relevant
proper functions is performed,
\[
\Gamma^{n,m,l}(\rho p_{1},\cdots,\rho p_{k})=\sum_{j=0}^{\infty}\rho^{j}\Gamma_{j}^{n,m,l}(p_{1},\cdots,p_{k})\,.
\]
Putting $\rho=0$ on both sides of this equation implies that $\Gamma_{0}^{n,m,l}$
does not depend on the momenta, in general the coefficient $\Gamma_{j}^{n,m,l}(p_{1},\cdots,p_{k})$,
will be a polynomial of degree $n+m+l$ in the momenta. 

In what follows, the field $\phi$ will be re-scaled as the field
$\rho$ in Appendix~\ref{sec:D-trick} \footnote{Appendix \ref{sec:D-trick} shows how the delta function trick works
for a scalar field theory.}, i.e.,
\begin{equation}
\phi=\sqrt{\lambda}\tilde{\phi}\label{eq:rescphi}
\end{equation}
this re-scaling makes the proper functions involving this field to
be also re-scaled. For example, the following term in the effective
action is rewritten as follows,
\[
\Gamma^{0,0,2}\phi\phi=\Gamma^{0,0,2}\lambda\tilde{\phi}\tilde{\phi}=\tilde{\Gamma}^{0,0,2}\tilde{\phi}\tilde{\phi}\quad\Rightarrow\quad\tilde{\Gamma}^{0,0,2}=\lambda\Gamma^{0,0,2}
\]
in general,
\[
\tilde{\Gamma}^{n,m,l}=\lambda^{l/2}\Gamma^{n,m,l}\,.
\]

To 0th order in $\rho$ the proper functions are constants independent
of the momenta\footnote{The resulting integrals appearing in r.h.s. of the flow equations
in this approximation are given in Appendix \ref{sec:C-diag} . This
graphs have a multiplicity which is computed in Appendix \ref{sec:G-mul}
.}. Thus, the proper functions appearing in the r.h.s. of the anomalous
dimensions $\gamma_{\psi}$, $\gamma_{A}$ and $\gamma_{\phi}$ are
constant. The calculation of $\gamma_{A}$, and of the first term
in $\gamma_{\psi}$ are the usual ones in QED. In this respect, it
is important to note, that since solutions of both the ERG equations
and the WT identities are considered then, the vacuum polarization
tensor $\Pi_{\mu\nu}$ appearing in the calculation of $\Gamma^{2,0,0}$
is a transversal one, i.e. satisfies the WT identity $k_{\mu}\Pi_{\mu\nu}(k)=0$.
In this respect it is also important to remark that in the forth equation
in Fig.~\ref{fig:erg} there is a cancellation between the first
diagram on the r.h.s. and $\gamma_{\psi}$, this is a consequence
of using dimensional regularization to compute the one loop integrals
and the WT identity ($Z_{1}=Z_{2}$) which relates the interaction
in $\Gamma^{1,2,0}$ and the kinetic term for the fermions appearing
in $\Gamma^{0,2,0}$. The second term in $\gamma_{\psi}$ does not
contribute when $\lambda\to0$, and the calculation of $\gamma_{\phi}$
only involves a fermion loop without $\gamma$ matrices in the vertices.
This leads to,
\begin{align*}
\gamma_{A} & =\frac{e^{2}}{6\pi^{2}}\;,\;\gamma_{\psi}=\frac{e^{2}}{16\pi^{2}}\;,\;\tilde{\gamma}_{\phi}=\frac{\lambda_{Y}^{2}}{4\pi^{2}}
\end{align*}
where the tilde in $\tilde{\gamma}_{\phi}$ indicates that this is
calculated using the re-scaled proper function $\tilde{\Gamma}^{0,0,2}$. 

\section{Long distance model for positronium interactions in QED}

It is convenient to define the following dimensionless parameters,
\begin{align*}
x_{1} & =\Gamma^{0,2,0}(0,\mu)\mu^{-1}=\frac{m}{\mu}\,,\\
x_{2} & =\lambda\Gamma^{0,0,2}(0,\mu)\mu^{-2}\,,\\
x_{3} & =\Gamma^{1,2,0}\mu^{-\epsilon/2}=e\mu^{-\epsilon/2}\,,\\
x_{4} & =\lambda^{1/2}\Gamma^{0,2,1}\mu^{-\epsilon/2}\,,\\
x_{5} & =\lambda^{2}\Gamma^{0,0,4}\mu^{-\epsilon}\,,\\
x_{6} & =\Gamma^{0,4,0}\mu^{2-\epsilon}\,,\\
x_{7} & =\lambda^{3/2}\Gamma^{0,0,3}\mu^{-1}\,.
\end{align*}
the definition of $x_{2},x_{4},x_{5}$ and $x_{7}$ has been done
in terms of the proper functions corresponding to the field $\tilde{\phi}$.
As mentioned in Section \ref{sec:The-effective-action}, the independent
parameters to be considered are $m,\tilde{m},e,\lambda_{Y},\lambda_{4},\lambda_{4f}$
and $\lambda_{3}$. 

\subsection{Discussion\label{subsec:Discussion}}

The scalar potential,
\[
V(\phi)=\mu^{2}x_{2}\phi^{2}+\mu\frac{x_{7}}{3!}\phi^{3}+\frac{x_{5}}{4!}\phi^{4}
\]
includes a $\phi^{3}$ term. This term makes the potential not symmetric
under $\phi\to-\phi$. This entails the appearance of two minima with
different values for the potential. How are these two minima related
to positronium? A natural way is to identify the lower minima with
the most stable state of positronium and the other minima to the next
stabler state. This interpretation is interesting because, introducing
higher powers of the field in the potential, would in principle allow
to describe additional excited states. In this respect, it is important
to remark that theoretically two different descriptions of positronium
are employed in relation to its instability. One that corresponds
to the, so to say, atomic decay of the energy levels of the positronium
atom and the other which comes from the decay of positronium by its
relativistic interaction with the electromagnetic field. These two
descriptions are considered and have non-trivial consequences. Indeed
the ground state of the positronium atom, called para-positronium
is less stable\footnote{This state is a scalar and can decay into two photons.}
than the first excited state called ortho-positronium\footnote{This state is a vector and can not decay into two photons.}.
The lifetimes and masses of these states are, 
\[
\tau_{pp}=0.12\,\text{{ns}}\;,\;m_{pp}=2m_{r}-6.8\,\text{{eV}},\;\tau_{op}=140\,\text{{ns}}\;,\;m_{op}=m_{pp}+0.001\,\text{{eV}}\,.
\]
Recalling the value of the rest electron mass $m_{r}=5.11\times10^{6}\,\text{{eV}}$shows
that $(m_{pp}-m_{op})/m_{r}\sim10^{-9}$. This value leads to a extremely
small coefficient for $x_{7}$ the coefficient of the $\phi^{3}$
interaction. So that for the phenomenological description of positronium
the effect of the $\phi^{3}$ term is negligible. However, in bound
states with higher separation between energy levels, terms such as
the $\phi^{3}$ in this case will have a non-negligible effect. Below
the case with no cubic term in the potential ($x_{7}=0$) is described,
however due to its relevance for the application of these ideas to
other bound states, the case with $x_{7}\neq0$ is considered in Appendix
H.

\subsection{The electron mass}

The electron mass in this model has two contributions: one from the
explicit mass term proportional to $x_{1}$, and the other generated
by a non-vanishing expectation value for the positronium scalar field
$\phi$ in the vacuum of this theory. This second contribution can
be expressed in terms of the parameters\footnote{For the case with the cubic term $x_{7}\neq0$, see Appendix H. }
$x_{2}$ and $x_{5}$, describing the scalar field potential, and
the Yukawa coupling $x_{4}$. This potential is:
\[
V(\phi)=\mu^{2}x_{2}\phi^{2}+\frac{x_{5}}{4!}\phi^{4}\,.
\]
The extrema of this potential are given by:
\[
\phi=0,\phi_{\mp}=\mp\mu\sqrt{-6\frac{x_{2}}{x_{5}}}\,
\]
where $\phi=0$ is a maximum. Thus, the electron mass is given by
$x_{1}\,\mu$ plus the Yukawa coupling $x_{4}$ times the vacuum expectation
value (v.e.v.) of the field $\phi$ at the true vacuum $\phi_{+}.$
Thus,
\begin{equation}
\frac{m_{e}}{\mu}=x_{1}+x_{4}\sqrt{-6\frac{x_{2}}{x_{5}}}\,.\label{eq:emass}
\end{equation}
It is worth noting that redefining the field $\phi$ as,
\[
\phi\to\phi'=\phi+\frac{x_{1}}{x_{4}}\mu
\]
then the relevant effective action terms change to,
\[
-x_{1}\mu\bar{\psi}\psi-x_{4}\phi\bar{\psi}\psi+V(\phi)\to-x_{4}\phi'\bar{\psi}\psi+V(\phi'-\frac{x_{1}}{x_{4}}\mu)
\]
thus the explicit mass term can be eliminated by a shift in the field
$\phi$. This shift does not alter the masses associated to each of
the minima and has no observable consequences. This can be explicitly
verified by maintaining the initial value of $m_{e}$ at the scale
$\mu=m_{r}$ and changing the value of $x_{1}(0)$.

\subsection{The positronium}

Regarding the positronium mass squared, this is given by the coefficient
of the expansion of the potential $V(\phi)$ around the vacuum $\phi_{\pm}$,
that is,
\[
m_{p}^{2}=\left.\frac{\partial^{2}V(\phi)}{\partial\phi^{2}}\right|_{\phi=\phi_{\pm}}=-2x_{2}\mu^{2}\,.
\]
This mass at the scale $\mu=m_{r}$(i.e. $t=0$) is $2m_{r}-b$, where
$b$ the binding energy of the ortho-positronium state, given by $b=6.8\,\text{{eV}}$,
\[
\left(2-\frac{b}{m_{r}}\right)^{2}=4-5\times10^{-5}=-2x_{2}(t=0)\,.
\]

\subsection{Positronium decay }

The positronium mean lifetime is $\tau\simeq0.12\,\text{{ns}}$. The
most important decay mode is into two photons. The first non-trivial
contribution is given by the following diagram,

\begin{figure}[H]
\includegraphics[scale=0.3]{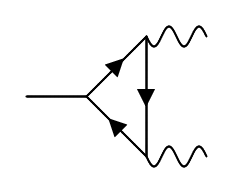}

\caption{The first contribution to positronium decay.}
\end{figure}

This diagram also appears in the fermion loop contribution to the
Higgs decay into two photons. Using the results in ref.~\cite{Marciano_2012}
\cite{Slominski1667828} leads to the following expression for the
positronium decay width,
\begin{equation}
\Gamma=\frac{\alpha^{2}}{256\pi^{3}}\frac{x_{4}^{2}}{2}16\frac{m_{e}^{2}}{m_{p}}|g(\beta)|^{2}\;,\;\beta=\frac{4m_{e}^{2}}{m_{P}^{2}}>1\label{eq:dw}
\end{equation}
where,
\[
g(\beta)=-2(1+(1-\beta)f(\beta))\;,\;f(\beta)=\arcsin(\beta^{-1/2})\,.
\]

\subsection{The $4$-fermion coupling}

Due to the delta function (\ref{eq:deltafipsi}), at long distances
the parameter $x_{6}$ is given in terms of $x_{2}$ and $x_{4}$
by:
\begin{equation}
x_{6}=\frac{x_{4}^{2}}{4x_{2}}\,.\label{eq:x6}
\end{equation}
This relation is employed to fix the initial values of $x_{6}$ at
long distances. 

\subsection{Initial conditions for the flow equations}

The initial conditions for the flow equations will be determined by
fixing the electron mass, its charge, the positronium mass, and its
decay width. Taking the reference scale $\mu_{0}=m_{r}$, where $m_{r}=0.51\,\text{{MeV}}$
is the electron mass at this scale, then $t=\log(\mu/\mu_{0})=0$
corresponds to $\mu=m_{r}$. The initial condition for the electron
charge is,
\begin{equation}
x_{3}(0)=-\sqrt{\frac{1}{137}}\,.\label{eq:charge}
\end{equation}
Regarding the electron mass, the initial condition is taken at its
rest mass $m_{r}$, which employing (\ref{eq:emass}) leads to,
\begin{equation}
1=x_{1}+x_{4}\sqrt{-6\frac{x_{2}}{x_{5}}}(t=0)\,.\label{eq:emass-mp}
\end{equation}
Regarding the positronium mass squared, this is given by the second
order coefficient in the expansion of the potential $V(\phi)$ around
the false vacuum $\phi_{+}$. This is given by,
\[
m_{p}^{2}=\left.\frac{\partial^{2}V(\phi)}{\partial\phi^{2}}\right|_{\phi=\phi_{+}}=-2x_{2}\mu^{2}\,.
\]
This mass at the scale $\mu=m_{r}$(i.e. $t=0$) is $2m_{r}-b$ where
$b$ the binding energy of the positronium ground state, given by
$b=6.8\,\text{{eV}},$ leading to,
\begin{equation}
4-3\times10^{-5}=-2x_{2}(t=0)\label{eq:mp-mp}
\end{equation}
Regarding the positronium decay width, eq.~(\ref{eq:dw}) will be
employed to obtain the initial value of $x_{4}$ at the scale $\mu=m_{r}$.
This leads to,
\begin{equation}
x_{4}^{2}=\Gamma\frac{m_{p}}{m_{e}^{2}}\frac{256\pi^{3}}{8\alpha^{2}|g(\beta)|^{2}}\label{eq:decayp}
\end{equation}
Thus, there are five equations~(\ref{eq:charge}),~(\ref{eq:emass-mp}),~(\ref{eq:mp-mp}),~(\ref{eq:decayp}),
and~(\ref{eq:x6}) for the five required initial conditions. 

\subsection{The flow equations and critical points}

The flow equations for the dimension-less parameters $x_{2},x_{3},x_{4},x_{5}$
and $x_{6}$ are, 
\begin{align*}
\dot{x}_{1} & =-x_{1}+\frac{x_{1}^{3}x_{6}}{4\pi^{2}}+\frac{2x_{1}x_{3}^{2}}{\pi^{2}}-\frac{x_{1}x_{4}^{2}}{2\pi^{2}}+\frac{x_{3}^{2}}{8\pi^{2}}\,,\\
\dot{x}_{2} & =-2x_{2}+\frac{3x_{1}^{2}x_{4}^{2}}{8\pi^{2}}+\frac{x_{2}x_{5}}{32\pi^{2}}+\frac{x_{4}^{2}}{4\pi^{2}}\,,\\
\dot{x}_{3} & =\frac{x_{3}^{3}}{12\pi^{2}}-\frac{x_{3}\text{\ensuremath{x_{4}^{2}}}}{48\pi^{2}}\,,\\
\dot{x}_{4} & =-\frac{9x_{1}^{2}x_{4}x_{6}}{8\pi^{2}}+\frac{x_{3}^{2}x_{4}}{6\pi^{2}}-\frac{x_{4}^{3}}{24\pi^{2}}\,,\\
\dot{x}_{5} & =\frac{x_{4}^{4}}{2\pi^{2}}+\frac{x_{4}^{2}x_{5}}{2\pi^{2}}-\frac{3x_{5}^{2}}{16\pi^{2}}\,,\\
\dot{x}_{6} & =2x_{6}-\frac{25\text{\ensuremath{x_{1}^{2}}}\text{\ensuremath{x_{6}^{2}}}}{16\pi^{2}}-\frac{x_{3}^{2}x_{6}}{6\pi^{2}}+\frac{5x_{4}^{2}x_{6}}{48\pi^{2}}\,.
\end{align*}
The critical points are obtained equating to zero all the beta functions.
The only non-trivial solution is,
\[
x_{1}=x_{2}=x_{3}=x_{4}=x_{5}=x_{6}=0\,.
\]
The flow equations are first order ordinary non-linear differential
equations, they can be solved numerically. The solutions with the
initial conditions given above are shown in the following plots.

\begin{figure}[H]
\begin{centering}
\includegraphics[scale=1.1]{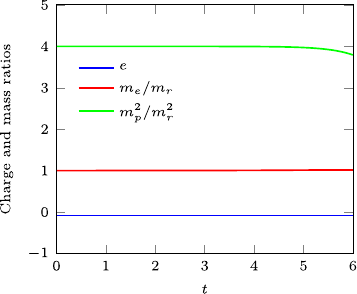}
\par\end{centering}
\caption{Masses and charge as a function of $t=\log(\mu/m_{r})$}
\end{figure}

\begin{figure}[H]
\begin{centering}
\includegraphics[scale=1.1]{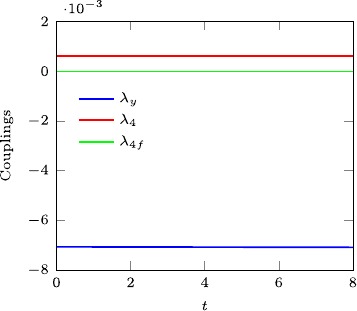}
\par\end{centering}
\caption{Couplings as a function of $t=\log(\mu/m_{r})$ }

\end{figure}
\begin{figure}[H]
\begin{centering}
\includegraphics[scale=1.1]{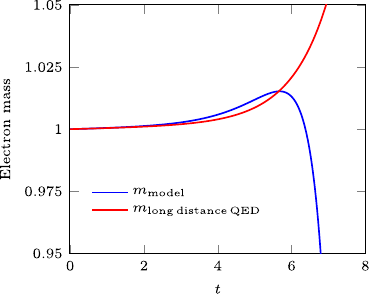}
\par\end{centering}
\caption{The electron mass in units of its rest mass for this model and long
distance QED without condensate.}
\end{figure}
\begin{figure}[H]
\begin{centering}
\includegraphics[scale=1.1]{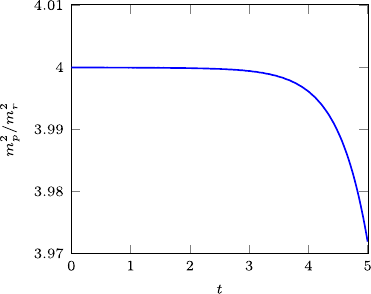}
\par\end{centering}
\caption{Positronium mass squared in units of $m_{r}^{2}$.}
\end{figure}

\section{Conclusions and outlook}

Conclusions and further research motivated by this work are summarized
in the series of remarks given below, 
\begin{itemize}
\item It has been shown that gauge theories can be consistently treated
using the dimensionally regulated ERG equations.
\item Solutions for both the ERG equations and the WT identities are considered
and obtained in a non-perturbative approximation scheme which consists
in a truncation and a low-momentum expansion.
\item This approach describes the long-distance physics of the positronium
bound state in QED, leading to an effective model consistent with
the available data.
\item It is worth noting that the delta function trick employed here is
quite general and can be used to describe more complex composite objects.
For example in describing multi-particle bound states. 
\item The implementation of this type of approach in the non-abelian case
is a natural and very interesting next step. 
\end{itemize}

\section*{Appendix A. Euclidean Dirac fermions.\label{sec:Appendix-A.-Gamma}}

Euclidean $4$-dimensional gamma matrices are considered. They are
defined by the following properties,
\begin{align*}
\{\gamma_{\mu},\gamma_{\nu}\} & =2\delta_{\mu\nu}\;,\;\gamma_{\mu}^{\dagger}=\gamma_{\mu}\\
\gamma_{5}=\gamma_{1}\gamma_{2}\gamma_{3}\gamma_{0} & =\gamma_{5}^{\dagger}\;,\;\gamma_{5}^{2}=1\;,\;\{\gamma_{5},\gamma_{\mu}\}=0\;\forall\mu
\end{align*}
\begin{comment}
thus,
\[
P_{\overset{L}{R}}=\frac{1\pm\gamma_{5}}{2}=P_{\overset{L}{R}}^{\dagger}
\]
and defining,
\[
\psi_{\overset{L}{R}}=\frac{1\pm\gamma_{5}}{2}\psi
\]
 then,
\[
\left(\bar{\psi}_{R}\psi_{L}\right)^{\dagger}=\bar{\psi}_{L}\psi_{R}\,.
\]
\end{comment}
\emph{The propagator in Euclidean space. }The action in Minkowski
space is given by,
\[
S_{M}=\int d_{M}^{4}x\,\bar{\psi\,}(i\gamma^{\mu}\partial_{\mu}-e\gamma^{\mu}A_{\mu}-m)\,\psi
\]
using that,
\[
\gamma_{4}=\gamma_{0}\;,\;\gamma_{i}=-i\gamma_{M}^{i}\;,\;x_{4}=ix_{0}\;,\;\partial_{4}=-i\partial_{0}\;,\;A_{4}=-iA_{0}
\]
leads to,
\[
S_{M}=i\int d_{E}^{4}x\,\bar{\psi\,}(\gamma_{\mu}\partial_{\mu}+ie\gamma_{\mu}A_{\mu}-m)\,\psi=i\,S
\]
where $S$ is the Euclidean action given by,
\[
S=\int d_{E}^{4}x\,\bar{\psi\,}(\gamma_{\mu}\partial_{\mu}+ie\gamma_{\mu}A_{\mu}-m)\,\psi\,.
\]
The factor $e^{iS_{M}}$ in the functional integral is therefore $e^{-S}$.
In momentum space the free part of $S$ is given by,
\[
S=\int d^{4}p\,\bar{\psi}(p)(i\gamma_{\mu}p_{\mu}-m)\psi(-p)\;\;,\;p_{\mu}=-i\partial_{\mu}
\]
\begin{comment}
defining the Fujikawa gamma matrices by,
\[
\gamma_{4}^{F}=i\gamma_{0}\;,\;\gamma_{i}^{F}=\gamma_{i}^{M}
\]
leads to,
\[
S_{M}=-i\int d_{E}^{4}x\,\bar{\psi}\,(-\gamma_{\mu}^{F}\partial_{\mu}-ie\gamma_{\mu}A_{\mu}-m)\,\psi=iS_{E}^{F}\Rightarrow S_{E}^{F}=\int d_{E}^{4}x\,\bar{\psi\,}(\gamma_{\mu}^{F}\partial_{\mu}+ie\gamma_{\mu}A_{\mu}+m)\,\psi\,.
\]
\end{comment}

\noindent and thus the Green function in momentum space,
\[
\frac{1}{-i\gamma_{\mu}^{E}p_{\mu}^{E}+m}=\frac{(i\gamma_{\nu}^{E}p_{\nu}^{E}+m)}{p_{E}^{2}+m^{2}}
\]
in coordinate space the propagator is,
\[
S(x-y)=\int\frac{d^{4}p}{(2\pi)^{4}}\,S(p)\,e^{-ip\cdot(x-y)}
\]
such that,
\[
(\gamma_{\mu}\partial_{\mu}+m)\,S(x-y)=\delta(x-y)\,.
\]

\section*{Appendix B. Integrals over momenta.\label{sec:Appendix-B.-Integrals}}

This section recalls results obtained in \cite{Trinchero_2022}. The
integral employed in this paper is,

\begin{align}
\int d^{n}p\,\left(\mu\frac{\partial}{\partial\mu}\theta(\frac{p^{2}}{\mu^{2}}-1)\right)\,\frac{f(p,q)}{p^{2}+m^{2}} & =\epsilon\int d^{d}p\mu^{\epsilon}\,\frac{f(p,q)}{p^{2}+m^{2}}+\nonumber \\
+\lim_{\epsilon\to0}\frac{S_{n-1}}{S_{d-1}}\int d^{d}p\mu^{\epsilon} & \frac{\mu^{2}}{p^{2}}\left(\left(p\left(\frac{f^{(1,0)}(p,q)}{m^{2}+p^{2}}-\frac{2pf(p,q)}{\left(m^{2}+p^{2}\right)^{2}}\right)\right.\right.\label{eq:intdtheta}\\
 & \left.\left.-2(\epsilon-1)\frac{f(p,q)}{p^{2}+m^{2}}\right)+\mathcal{O}\left(\frac{\mu^{4}}{p^{4}}\right)\cdots\right)
\end{align}
where the scale parameter $\mu$ is related to $\epsilon$ by, 
\[
\mu^{2}\propto\mu_{0}^{2}\,e^{-\frac{1}{|\epsilon|}}
\]
and $\mu_{0}$ is a reference scale. 

\section*{Appendix C. Diagrams to be computed\label{sec:C-diag}}

It is convenient to define the following dimensionless parameters
which arise from the rescaling (\ref{eq:rescphi}) of the field $\phi$,
\begin{align*}
x_{1} & =\Gamma^{0,2,0}(0,\mu)\,\mu^{-1}\,,\\
x_{2} & =\lambda\Gamma^{0,0,2}\,\mu^{-2}=\tilde{\Gamma}^{0,0,2}(0,\mu)\,\mu^{-2}\,,\\
x_{3} & =\Gamma^{1,2,0}\,\mu^{-\epsilon/2}\,,\\
\tilde{x}_{4} & =\lambda^{1/2}\,\Gamma^{0,2,1}\,\mu^{-\epsilon/2}=\tilde{\Gamma}^{0,2,1}\,\mu^{-\epsilon/2}\,,\\
x_{5} & =\lambda^{2}\,\Gamma^{0,0,4}\,\mu^{-\epsilon}=\tilde{\Gamma}^{0,0,4}\,\mu^{-\epsilon}\,,\\
x_{6} & =\Gamma^{0,4,0}\,\mu^{2-\epsilon}\,,\\
x_{7} & =\lambda^{3/2}\,\Gamma^{0,0,3}\,\mu^{-1}=\tilde{\Gamma}^{0,0,3}\,\mu^{-1}\,.
\end{align*}
The following diagrams appear on the r.h.s. of the flow equations.
In the r.h.s of the equations below $M_{n,2m,l}$ denotes the multiplicity
of the corresponding contribution which are described and given in
Appendix \ref{sec:G-mul}.\vspace{0.5cm}

\noindent Contribution to the r.h.s. of the ERG equation for $\Gamma^{2,0,0}$:

\vspace{0.25cm}

\begin{flushleft}
\includegraphics[bb=0bp 23.92523bp 173.7868bp 64bp,scale=0.25]{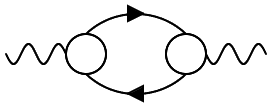}$=-\frac{M_{2,0,0}}{2!}\,x_{3}^{2}\,\mu^{\epsilon}\epsilon\int\frac{d^{d}p}{(2\pi)^{d}}\,\text{{Tr}}\left[\gamma_{\mu}\frac{1}{i\not p-m}\gamma_{\mu}\frac{1}{i(\not p+\not\not q)-m}\right]\overset{\epsilon\to0}{=}\frac{2}{3}\,\frac{x_{3}^{2}}{\pi^{2}}\,(\delta_{\mu\nu}q^{2}-q_{\mu}q_{\nu})\,.$
\par\end{flushleft}

\vspace{0.5cm}

\noindent Contribution to the r.h.s. of the ERG equation for $\Gamma^{0,2,0}$:

\vspace{0.25cm}

\begin{flushleft}
\includegraphics[bb=0bp 25.67626bp 174.428bp 83bp,scale=0.25]{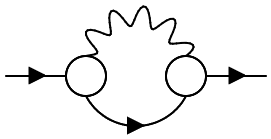}$=\frac{M_{0,2,0}^{(1)}}{2!}\,x_{3}^{2}\,\mu^{\epsilon}\epsilon\int\frac{d^{d}p}{(2\pi)^{d}}\,\frac{1}{p^{2}}\gamma_{\mu}\left(\frac{(i\not p+m)}{p^{2}+m^{2}}\right)\gamma^{\mu}\overset{\epsilon\to0}{=}-2\mu\,\frac{x_{1}x_{3}^{2}}{\pi^{2}}\,,$
\par\end{flushleft}

\begin{flushleft}
\includegraphics[bb=0bp 25.87288bp 155.7759bp 71bp,scale=0.25]{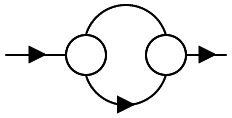}$=\frac{M_{0,2,0}^{(2)}}{2!}\,x_{4}^{2}\,\mu^{\epsilon}\epsilon\int\frac{d^{d}p}{(2\pi)^{d}}\,\left(\frac{(i\not p+m)}{p^{2}+m^{2}}\right)\frac{1}{p^{2}+m_{\phi}^{2}}$$\overset{\epsilon\to0}{=}\mu\,\frac{x_{1}\tilde{x}_{4}^{2}}{\pi^{2}}\,,$
\par\end{flushleft}

\begin{flushleft}
\includegraphics[bb=0bp 13.16393bp 131.5464bp 73bp,scale=0.25]{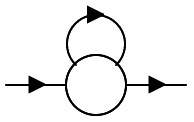}$=-M_{0,2,0}^{(3)}\,x_{6}\,\mu^{\epsilon-2}\epsilon\int\frac{d^{d}p}{(2\pi)^{d}}\,\frac{(i\not p+m)}{p^{2}+m^{2}}\overset{\epsilon\to0}{=}-\frac{\mu}{2}\,\frac{x_{1}^{3}x_{6}}{\pi^{2}}\,.$\vspace{0.5cm}
\par\end{flushleft}

\noindent Contribution to the r.h.s. of the ERG equation for $\Gamma^{0,0,2}$:

\vspace{0.25cm}

\includegraphics[bb=0bp 25.2bp 179.562bp 66bp,scale=0.25]{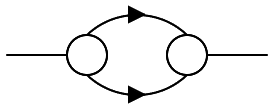}$=-\frac{M_{0,0,2}^{(1)}}{2!}\,x_{4}^{2}\,\mu^{\epsilon}\epsilon\int\frac{d^{d}p}{(2\pi)^{d}}\,\text{{Tr}}\left[\left(\frac{(i\not p+m)}{p^{2}+m^{2}}\right)^{2}\right]\overset{\epsilon\to0}{=}-3\,\lambda^{-1}\mu^{2}\,\frac{x_{1}^{2}\tilde{x}_{4}^{2}}{\pi^{2}}\,,$

\includegraphics[bb=0bp 9.62963bp 126bp 65bp,scale=0.25]{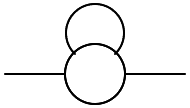}$=M_{0,0,2}^{(2)}\,\mu^{\epsilon}\epsilon\int\frac{d^{d}p}{(2\pi)^{d}}\,\frac{1}{p^{2}+\tilde{m}_{\phi}^{2}}\overset{\epsilon\to0}{=}\mu^{2}\frac{x_{2}x_{5}}{8\pi^{2}}\,.$\vspace{0.5cm}

\noindent Contribution to the r.h.s. of the ERG equation for $\Gamma^{1,2,0}$:

\vspace{0.25cm}

\includegraphics[bb=0bp 2.40884bp 179.7794bp 109bp,scale=0.25]{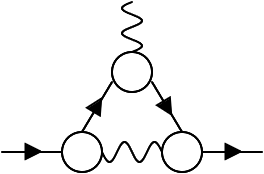}$=\frac{M_{1,2,0}^{(1)}}{3!}\,x_{3}^{3}\,\mu^{\epsilon}\epsilon\int\frac{d^{d}p}{(2\pi)^{d}}\,\gamma_{\rho}\frac{(i\not p+m)}{p^{2}+m^{2}}\gamma_{\mu}\frac{(i\not p+m)}{p^{2}+m^{2}}\gamma_{\nu}\frac{g^{\nu\rho}}{p^{2}}\overset{\epsilon\to0}{=}\frac{1}{4}\,\gamma_{\mu}\,\frac{x_{3}^{3}}{\pi^{2}}\,,$

\includegraphics[bb=0bp 2.4bp 179.7794bp 108bp,scale=0.25]{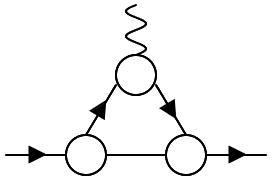}$=\frac{M_{1,2,0}^{(2)}}{3!}\,x_{3}x_{4}^{2}\,\mu^{\epsilon}\epsilon\int\frac{d^{d}p}{(2\pi)^{d}}\,\frac{(i\not p+m)}{p^{2}+m^{2}}\frac{\gamma_{\mu}}{p^{2}+\tilde{m}^{2}}\frac{(i\not p+m)}{p^{2}+m^{2}}\overset{\epsilon\to0}{=}-\frac{1}{6}\,\gamma_{\mu}\,\frac{x_{3}x_{4}^{2}}{\pi^{2}}\,,$

\includegraphics[bb=0bp 9.6bp 132bp 126bp,scale=0.25]{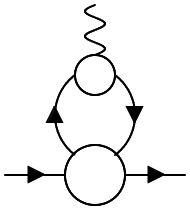}$=-\frac{M_{1,2,0}^{(3)}}{2!}\,x_{3}x_{6}\,\mu^{\epsilon-2}\epsilon\int\frac{d^{d}p}{(2\pi)^{d}}\,\frac{(i\not p+m)}{p^{2}+m^{2}}\gamma_{\mu}\frac{(i\not p+m)}{p^{2}+m^{2}}\overset{\epsilon\to0}{=}0\,.$

\vspace{0.5cm}

\noindent Contribution to the r.h.s. of the ERG equation for $\Gamma^{0,2,1}$:

\vspace{0.25cm}

\includegraphics[bb=0bp 3.6bp 168bp 90bp,scale=0.25]{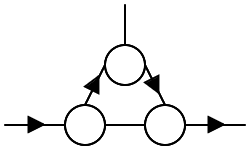}$=\frac{M_{0,2,1}^{(1)}}{3!}\,x_{4}^{3}\,\mu^{\epsilon}\epsilon\int\frac{d^{d}p}{(2\pi)^{d}}\,\frac{(i\not p+m)}{p^{2}+m^{2}}\frac{1}{p^{2}+m_{\phi}^{2}}\frac{(i\not p+m)}{p^{2}+m^{2}}\overset{\epsilon\to0}{=}0\,,$

\includegraphics[bb=0bp 2.4bp 174.428bp 102bp,scale=0.25]{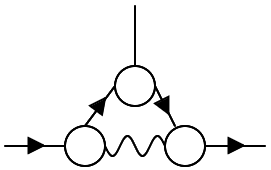}$=\frac{M_{0,2,1}^{(2)}}{3!}x_{3}^{2}x_{4}\mu^{\epsilon}\epsilon\int\frac{d^{d}p}{(2\pi)^{d}}\,\gamma_{\mu}\frac{(i\not p+m)}{p^{2}+m^{2}}\frac{(i\not p+m)}{p^{2}+m^{2}}\frac{1}{p^{2}}\gamma^{\mu}\overset{\epsilon\to0}{=}-\frac{2}{3}\,\frac{x_{3}^{2}x_{4}}{\pi^{2}}\,,$

\includegraphics[bb=0bp 9.58242bp 131.7708bp 109bp,scale=0.25]{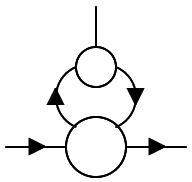}$=-\frac{M_{0,2,1}^{(2)}}{2!}x_{6}x_{4}\mu^{\epsilon-2}\epsilon\int\frac{d^{d}p}{(2\pi)^{d}}\,\frac{(i\not p+m)}{p^{2}+m^{2}}\frac{(i\not p+m)}{p^{2}+m^{2}}\overset{\epsilon\to0}{=}-9\,\frac{x_{1}^{2}x_{4}x_{6}}{\pi^{2}}\,,$

\includegraphics[bb=0bp 3.59211bp 143.7736bp 91bp,scale=0.25]{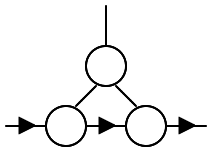}$=\frac{M_{0,2,1}^{(2)}}{3!}x_{7}x_{4}^{2}\mu^{\epsilon-1}\epsilon\int\frac{d^{d}p}{(2\pi)^{d}}\,\frac{(i\not p+m)}{p^{2}+m^{2}}\frac{1}{p^{2}+m^{2}}\overset{\epsilon\to0}{=}0\,.$

\vspace{0.5cm}
\noindent Contribution to the r.h.s. of the ERG equation for $\Gamma^{0,0,3}$:

\vspace{0.25cm}

\includegraphics[bb=0bp 4.8bp 144bp 90bp,scale=0.25]{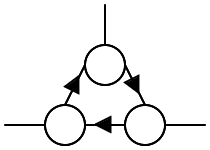}$=-\frac{M_{0,0,3}^{(1)}}{3!}\,x_{4}^{3}\,\mu^{\epsilon-1}\epsilon\int\frac{d^{d}p}{(2\pi)^{d}}\,\text{{Tr}}\left(\frac{-(i\not p+m)}{p^{2}+m^{2}}\right)^{3}\overset{\epsilon\to0}{=}-6\,\frac{x_{1}x_{4}^{3}}{\pi^{2}}\,,$

\vspace{0.5cm}

\noindent Contribution to the r.h.s. of the ERG equation for $\Gamma^{0,0,4}$:

\vspace{0.25cm}

\includegraphics[bb=0bp 24.06015bp 168.2353bp 80bp,scale=0.25]{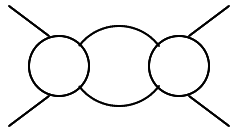}$=\frac{M_{0,0,4}^{(1)}}{2!}\,x_{5}^{2}\,\mu^{2\epsilon}\epsilon\int\frac{d^{d}p}{(2\pi)^{d}}\,\left(\frac{1}{p^{2}+m_{\phi}^{2}}\right)^{2}\overset{\epsilon\to0}{=}\frac{3}{8}\,\frac{x_{5}^{2}}{\pi^{2}}\,,$

\includegraphics[bb=0bp 62.9167bp 180.4428bp 151bp,scale=0.25]{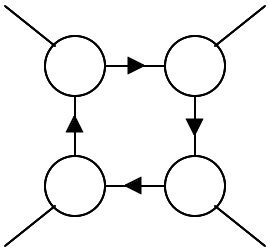}$=-\frac{M_{0,0,4}^{(2)}}{4!}\,x_{4}^{4}\,\mu^{2\epsilon}\epsilon\int\frac{d^{d}p}{(2\pi)^{d}}\,\text{{Tr}}\left[\left(\frac{(i\not p+m)}{p^{2}+m^{2}}\right)^{4}\right]\overset{\epsilon\to0}{=}-16\,\frac{x_{4}^{4}}{\pi^{2}}\,.$\vspace{0.5cm}

\noindent Contribution to the r.h.s. of the ERG equation for $\Gamma^{0,4,0}$\label{Gamma-040}:

\vspace{0.25cm}

\includegraphics[bb=0bp 25.26316bp 167.7586bp 80bp,scale=0.25]{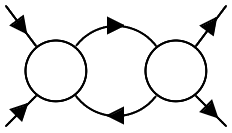}$=\frac{M_{0,4,0}^{(1)}}{2!}\,x_{6}^{2}\,\mu^{2\epsilon-4}\epsilon\int\frac{d^{d}p}{(2\pi)^{d}}\,\left(\frac{(i\not p+m)}{p^{2}+m^{2}}\right)^{2}\overset{\epsilon\to0}{=}\frac{25}{2}\,\frac{x_{1}^{2}x_{6}^{2}}{\pi^{2}}\,,$

\includegraphics[bb=0bp 57bp 168bp 138bp,scale=0.25]{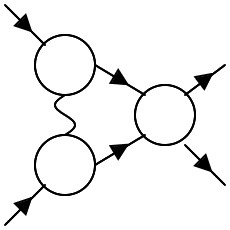}$=\frac{M_{0,4,0}^{(2)}}{3!}\,x_{3}^{2}x_{6}\,\mu^{2\epsilon-2}\epsilon\int\frac{d^{d}p}{(2\pi)^{d}}\,\gamma_{\mu}\frac{(i\not p+m)}{p^{2}+m^{2}}\frac{(i\not p+m)}{p^{2}+m^{2}}\frac{1}{p^{2}}\gamma^{\mu}\overset{\epsilon\to0}{=}-\frac{5}{3}\,\frac{x_{3}^{2}x_{6}}{\pi^{2}}\,,$

\includegraphics[bb=0bp 57bp 168bp 168bp,scale=0.25]{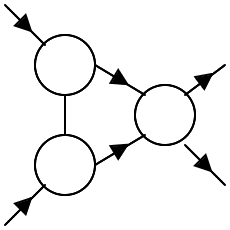}$=\frac{M_{0,4,0}^{(3)}}{3!}\,x_{4}^{2}x_{6}\,\mu^{2\epsilon-2}\epsilon\int\frac{d^{d}p}{(2\pi)^{d}}\,\frac{(i\not p+m)}{p^{2}+m^{2}}\frac{1}{p^{2}+m_{\phi}^{2}}\frac{(i\not p+m)}{p^{2}+m^{2}}\overset{\epsilon\to0}{=}0\,.$

\section*{Appendix D. The delta approximant trick\label{sec:D-trick}}

Consider the theory,
\begin{align}
Z & =\int\mathcal{D}\phi\,e^{-\left(\int\frac{1}{2}\phi(-\boxempty+m^{2})\phi+\frac{g}{2M^{2}}\phi^{2}(-\boxempty+M^{2})\phi^{2}\right)}\label{eq:z}\\
 & =\int\mathcal{D}\phi\mathcal{D}\rho\,e^{-\left(\int\frac{1}{2}\phi(-\boxempty+m^{2})\phi+\frac{g}{2}\rho(-\boxempty+M^{2})\rho\right)}\,\delta(M\rho-\phi^{2})\,.\nonumber 
\end{align}
Next, the following approximant of the delta function is employed,
\[
\delta(M\rho-\sqrt{g}\phi^{2})=\lim_{\lambda\to0}\frac{1}{\sqrt{2\lambda\pi}}\,e^{-\frac{1}{2\lambda}\int(M\rho-\sqrt{g}\phi^{2})^{2}}
\]
leading to,
\begin{align*}
Z & =\lim_{\lambda\to0}\frac{1}{\sqrt{2\lambda\pi}}\int\mathcal{D}\phi\mathcal{D}\rho\,e^{-\left(\int\frac{1}{2}\phi(-\boxempty+m^{2})\phi+\frac{1}{2}\rho(-\boxempty+M^{2}(1+\frac{1}{\lambda}))\rho+\frac{g}{2\lambda}\phi^{4}-\frac{\sqrt{g}}{\lambda}M\rho\phi^{2}\right)}\\
 & =\lim_{\lambda\to0}\frac{1}{\sqrt{2\lambda\pi}}\int\mathcal{D}\phi\mathcal{D}\tilde{\rho}\,e^{-\left(\int\frac{1}{2}\phi(-\boxempty+m^{2})\phi+\frac{1}{2}\lambda\tilde{\rho}(-\boxempty+M^{2}(1+\frac{1}{\lambda}))\tilde{\rho}+\frac{g}{2\lambda}\phi^{4}-\sqrt{\frac{g}{\lambda}}M\tilde{\rho}\phi^{2}\right)}
\end{align*}
where the second line is obtained by making the change of variables
$\rho=\sqrt{\lambda}\tilde{\rho}.$ First it is shown, how the first
expression for $Z$ is obtained by integrating out the field $\rho$.
Integrating out $\rho$ leads to,
\[
Z=\lim_{\lambda\to0}\frac{1}{\sqrt{\lambda}}\int\mathcal{D}\phi\,e^{-\left(\int\frac{1}{2}\phi(-\boxempty+m^{2})\phi+\frac{g}{2}\phi^{2}\frac{\left(-\boxempty+M^{2}\right)}{M^{2}(\lambda+1)-\boxempty\lambda}\phi^{2}\right)}
\]
which in the limit coincides with the starting expression for $Z$.

In what follows, the diagram corresponding to the the 1-loop correction
to the 2-point function for the field $\phi$ is considered. Without
introducing the delta function this is given by,
\begin{align*}
\Gamma_{2} & =-g\frac{12}{2M^{2}}\int\frac{d^{d}p}{(2\pi)^{d}}\frac{p^{2}+M^{2}}{p^{2}+m^{2}}=-\frac{6}{M^{2}}\int\frac{d^{d}p}{(2\pi)^{d}}\frac{p^{2}+m^{2}+M^{2}-m^{2}}{p^{2}+m^{2}}\\
 & =-\frac{6}{M^{2}}\int\frac{d^{d}p}{(2\pi)^{d}}\frac{M^{2}-m^{2}}{p^{2}+m^{2}}=6\frac{M^{2}-m^{2}}{M^{2}}\frac{m^{2}}{8\pi^{2}\epsilon}=6\frac{m^{2}}{8\pi^{2}\epsilon}\left(1-\frac{m^{2}}{M^{2}}\right)
\end{align*}
where the multiplicity factor $12$ is the one corresponding to this
tadpole diagram. On the other hand, introducing the delta function,
there are two diagrams contributing to the 1-loop correction to the
2-point proper $\phi$ function, which lead to,

\vspace{0.25cm}

\includegraphics[bb=0bp 24bp 511bp 97bp,scale=0.35]{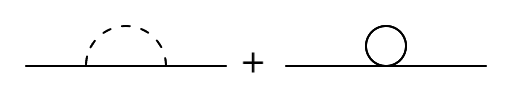}$=\frac{6g}{\lambda}M^{2}\int\frac{d^{d}p}{(2\pi)^{d}}\frac{1}{p^{2}+m^{2}}\frac{1}{\lambda(p^{2}+M^{2}(1+\frac{1}{\lambda}))}-\frac{12g}{2\lambda}\int\frac{d^{d}p}{(2\pi)^{d}}\frac{1}{p^{2}+m^{2}}$

\vspace{0.25cm}

\noindent using that,
\[
\frac{1}{\lambda(p^{2}+M^{2}(1+\frac{1}{\lambda}))}=\frac{1}{M^{2}}-\frac{\lambda(p^{2}+M^{2})}{M^{4}}
\]
leads to,

\vspace{0.25cm}

\noindent\includegraphics[bb=0bp 24bp 511bp 97bp,scale=0.35]{2-point-proper-fi}$=\frac{6g}{\lambda}M^{2}\int\frac{d^{d}p}{(2\pi)^{d}}\frac{1}{p^{2}+m^{2}}\left(\frac{1}{M^{2}}-\frac{\lambda(p^{2}+M^{2})}{M^{4}}\right)-\frac{12g}{2\lambda}\int\frac{d^{d}p}{(2\pi)^{d}}\frac{1}{p^{2}+m^{2}}$

\noindent$=\frac{6g}{\lambda}\left(\int\frac{d^{d}p}{(2\pi)^{d}}\frac{1}{p^{2}+m^{2}}-\frac{\lambda}{M^{2}}\frac{(p^{2}+M^{2})}{p^{2}+m^{2}}\right)-\frac{12g}{2\lambda}\int\frac{d^{d}p}{(2\pi)^{d}}\frac{1}{p^{2}+m^{2}}=-\frac{6g}{M^{2}}\int\frac{d^{d}p}{(2\pi)^{d}}\frac{(p^{2}+M^{2})}{p^{2}+m^{2}}\,.$

\vspace{0.25cm}

\noindent This result coincides with the expression appearing in the
computation without employing the delta function.

\section*{Appendix E. The dimensionally regularized ERG version of the Wilson-Fisher
fixed point.\label{fi4}}

The ERG equations for the $\Gamma_{2}$ and $\Gamma_{4}$ in a $\phi$
scalar field theory with symmetry under $\phi\to-\phi$ are,
\begin{align}
\mu\frac{\partial}{\partial\mu}\Gamma_{2}(p_{1};\mu) & =\left(\gamma+p_{1}\cdot\frac{\partial}{\partial p_{1}}\right)\,\Gamma_{2}(p_{1};\mu)+\nonumber \\
 & +\frac{\mu^{-\epsilon}}{2}\int\frac{d^{n}p}{(2\pi)^{n}}\left(\mu\frac{\partial}{\partial\mu}\Delta_{2}(p;\mu)\right)\,\Gamma_{4}(p,-p,p_{1},-p_{1};\mu)\label{eq:2p-full}
\end{align}
\begin{align}
\mu\frac{\partial}{\partial\mu}\Gamma_{4}(p_{1},p_{2},p_{3},p_{4};\mu) & =\left(2\gamma+\sum_{i=1}^{3}p_{i}\cdot\frac{\partial}{\partial p_{i}}\right)\,\Gamma_{4}(p_{1},p_{2},p_{3},p_{4};\mu)\nonumber \\
+\frac{\mu^{-\epsilon}}{2}\int\frac{d^{n}p}{(2\pi)^{n}} & \left(\mu\frac{\partial}{\partial\mu}\Delta_{2}(p;\mu)\right)\,\Gamma_{6}(p,-p,p_{1},p_{2},p_{3},p_{4};\mu)\nonumber \\
-\frac{\mu^{-\epsilon}}{2}\int\frac{d^{n}p}{(2\pi)^{n}}\frac{d^{n}p'}{(2\pi)^{n}} & \left(\mu\frac{\partial}{\partial\mu}\Delta_{2}(p;\mu)\Delta_{2}(p';\mu)\right)\nonumber \\
\times\left(\Gamma_{4}(p,-p',p_{1},p_{2};\mu)\right. & \,\Gamma_{4}(p',-p,p_{3},p_{4};\mu)\,\delta^{(n)}(p-p'+p_{1}+p_{2})\nonumber \\
+\Gamma_{4}(p,-p',p_{1},p_{3};\mu) & \,\Gamma_{4}(p',-p,p_{2},p_{4};\mu)\,\delta^{(n)}(p-p'+p_{1}+p_{3})\nonumber \\
+\Gamma_{4}(p,-p',p_{1},p_{4};\mu) & \left.\Gamma_{4}(p',-p,p_{2},p_{3};\mu)\,\delta^{(n)}(p-p'+p_{1}+p_{4})\right)\,.\label{eq:4p-full}
\end{align}
These equations are considered assuming a truncation where $\Gamma_{n}=0\;,\forall n>4$.
In addition, the proper functions $\Gamma_{2}$ and $\Gamma_{4}$
are assumed to be momentum independent\footnote{This assumption is equivalent to keeping only the first term in a
low momentum expansion of these proper functions. }. The approximated ERG equations are,
\begin{align*}
\mu\frac{\partial}{\partial\mu}\Gamma_{4}^{(0)}(\mu) & =-\frac{3}{2}\,\left(\Gamma_{4}^{2}(\mu)\mu^{-\epsilon}\right)\,\epsilon\int\frac{d^{d}p}{(2\pi)^{d}}\,\left(\frac{1}{p^{2}+m^{2}+\Sigma_{2}(\mu)}\right)^{2}\\
 & =-\frac{3}{(4\pi)^{2}}\left(\Gamma_{4}^{2}(\mu)\mu^{-\epsilon}\right)\,,
\end{align*}
\begin{align}
\mu\frac{\partial}{\partial\mu}\Gamma_{2}(\mu) & =\frac{1}{2}\,\left(\Gamma_{4}(\mu)\mu^{-\epsilon}\right)\,\epsilon\int\frac{d^{d}p}{(2\pi)^{d}}\,\left(\frac{1}{p^{2}+m^{2}+\Sigma_{2}(\mu)}\right)\nonumber \\
 & =-\frac{1}{2}\,\Gamma_{4}(\mu)\mu^{-\epsilon}\,\frac{m^{2}+\Sigma_{2}(\mu)}{8\pi^{2}}\,.\label{eq:dmug2}
\end{align}
It is noted that as explained in Weinberg's paper p. 43, the complete
proper two point function in this approximation is given by\footnote{This amounts diagrammatically to represent the r.h.s. of the ERG equations
with graphs that do not involve $2$-point proper functions. },
\[
\frac{1}{p^{2}+m^{2}+\Sigma_{2}(\mu)}
\]
where $m^{2}$ is the square mass without quantum corrections and
$\Sigma_{2}(\mu)$ is the quantum correction. Defining the dimensionless
couplings,
\[
x_{4}=\Gamma_{4}(\mu)\mu^{-\epsilon}\quad,\;x_{2}=\Sigma_{2}(\mu)\mu^{-2}
\]
then the corresponding beta functions are given by,
\begin{align*}
\mu\frac{\partial}{\partial\mu}x_{4} & =-\epsilon x_{4}+\mu^{-\epsilon}\mu\frac{\partial}{\partial\mu}\Gamma_{4}(\mu)\\
 & =-\epsilon x_{4}-\mu^{-\epsilon}\frac{3}{(4\pi)^{2}}\left(\Gamma_{4}^{2}(\mu)\mu^{-\epsilon}\right)=-\epsilon x_{4}-x_{4}^{2}\frac{3}{(4\pi)^{2}}\,,\\
\mu\frac{\partial}{\partial\mu}x_{2} & =-2x_{2}+\mu^{-2}\mu\frac{\partial}{\partial\mu}\Gamma_{2}(\mu)\\
 & \overset{(\ref{eq:dmug2})}{=}-2x_{2}-\frac{1}{2}x_{4}\frac{(m^{2}+\Sigma_{2}(\mu))\mu^{-2}}{8\pi^{2}}=-2x_{2}-\frac{1}{2}x_{4}\frac{(m^{2}/\mu^{2}+x_{2})}{8\pi^{2}}\,.
\end{align*}
The fixed points of this flow are the zeros of the beta functions,
thus they are given by,
\begin{align*}
-\epsilon x_{4}^{*}-x_{4}^{*2}\frac{3}{(4\pi)^{2}} & =0\Rightarrow\left\{ \begin{array}{c}
x_{4}^{*}=0\\
x_{4}^{*}=-\frac{(4\pi)^{2}}{3}\epsilon
\end{array}\right.\\
-2x_{2}^{*}-\frac{1}{2}x_{4}^{*}\frac{(m^{2}/\mu^{2}+x_{2}^{*})}{8\pi^{2}} & =0\Rightarrow x_{2}^{*}=\left\{ \begin{array}{c}
x_{2}^{*}=0\\
x_{2}^{*}=\frac{\epsilon}{6}\frac{m^{2}}{\mu^{2}}\frac{1}{\left(1+\frac{\epsilon}{6}\right)}
\end{array}\right.
\end{align*}
Next the linearized ERG equations are solved. These equations are
of the form:
\[
\mu\frac{\partial}{\partial\mu}x_{i}=f_{i}(\{x\})\;\;,\;i=2,4\,.
\]
Linearization consists in keeping only the linear term of the power
series expansion of the functions $f_{i}$ around the fixed points
$x^{*}$. This corresponds to the replacement,
\[
f_{i}(\{x\})\rightsquigarrow\left.\frac{\partial f_{i}(\{x\})}{\partial x_{j}}\right|_{x=x^{*}}\,\delta x_{j}=M_{ij}\,\delta x_{j}
\]
in this case, the matrix $M$ for the non-trivial fixed point, the
Wilson-Fisher one, is given by,
\[
M=\left(\begin{array}{cc}
-2-\frac{x_{4}^{*}}{(4\pi)^{2}} & \,\frac{(m^{2}/\mu^{2}+x_{2}^{*})}{(4\pi)^{2}}\\
0 & -\epsilon-\frac{6\,x_{4}^{*}}{(4\pi)^{2}}
\end{array}\right)=\left(\begin{array}{cc}
-2+ & \frac{\epsilon}{3}\,\frac{m^{2}}{(4\pi)^{2}\mu^{2}}\,\left(\frac{1}{1+\epsilon/6}\right)\\
0 & \epsilon
\end{array}\right)\,.
\]
The eigenvalues $\lambda_{i}$ of the matrix $M$ are related to the
critical exponents. The critical exponent $\nu,$ which describes
the behavior of the two-point function for long distances near the
critical point which is given by,
\[
G\sim\frac{e^{-r/\xi}}{r^{\frac{d-1}{2}}}\;\;,\;\;\xi\sim|t-t_{c}|^{-\nu}
\]
 is related to the eigenvalue of $M$ corresponding to $\Gamma_{2}$,
by,
\[
\nu=-\frac{1}{\lambda_{2}}=\frac{-1}{\frac{\epsilon}{3}-2}\overset{\epsilon\to1}{=}0.6
\]
this value is in reasonable agreement with the value $\nu=0.63$ for
the Ising model for $d=3$ (i.e. $\epsilon=1$). 

It is worth noting that this result for the critical exponents agrees
with the well known computed with a hard cut-off\cite{WF-PhysRevLett.28.240}.
This is quite remarkable because the location of critical points and
beta functions, are completely different. Indeed $x_{4}^{*}$ is positive
for usual approach and negative with this approach, also the beta
functions are different, however universal quantities such as the
critical exponents are the same.

\section*{Appendix F. Derivation of ERG equations.\label{Derivation}}

\noindent The generating functional $Z[J]$ of correlators is given
by,
\[
Z[J]=\int\mathcal{D}\phi\,e^{-S[\phi,t]+\int_{p}J_{-p}\phi_{p}}
\]
where,
\[
S[\phi,t]=\sum_{n}\frac{1}{n!}\int_{p_{1},\cdots,p_{n}}u_{n}(p_{1},\cdots,p_{n},t)\phi(p_{1})\cdots\phi(p_{n})\quad,\;t=\ln\frac{\mu}{\mu_{0}}
\]
where the dimensionless variable $t=\ln\mu/\mu_{0}$ is related to
the momentum scale $\mu$ which separates high and low momentum degrees
of freedom. The only $u_{n}$ function that will be assumed to be
depending on $t$ is the $u_{2}$, as follows,
\begin{equation}
u_{2}(p_{1},p_{2},t)=\tilde{u}_{2}(p_{1},p_{2})+G_{>}^{-1}(p_{1},t)\delta(p_{1}+p_{2})\;,\;G_{>}(p_{1},t)=\frac{\theta_{\epsilon}(|p|-\mu)}{p^{2}+m^{2}}\,.\label{eq:u2t}
\end{equation}
Thinking in terms of graphs where the internal lines are given by
the inverse of the second factor in the expression of $u_{2}(p_{1},p_{2},t)$,
this amounts to restrict the momenta of the internal lines to be larger
than the scale $\mu$, that means that high momentum degrees of freedom
are integrated out(the subindex $>$ on the propagator $G$ has been
included to make this explicit). Then,
\begin{align*}
S[\phi,t] & =\frac{1}{2}\int_{p_{1},p_{2}}G_{>}^{-1}(p_{1},t)\delta(p_{1}+p_{2})\phi(p_{1})\phi(p_{2})+\tilde{S}[\phi]\\
 & =S_{2}^{>}[\phi,t]+\tilde{S}[\phi]\,.
\end{align*}
The derivative with respect to $t$ of the generating functional of
connected correlators $W[J]=\ln Z[J]$ is given by, 
\begin{equation}
\partial_{t}W[J]=-\frac{1}{Z[J]}\int\mathcal{D}\phi\,e^{-S[\phi,t]+\int_{p}J_{-p}\phi_{p}}\,\partial_{t}\left(\frac{1}{2}\int_{p,q}\phi(p)G_{>}^{-1}(p,q,t)\phi(q)\right)\label{eq:dtw}
\end{equation}
where,
\[
G_{>}^{-1}(p,q,t)=G_{>}^{-1}(p,t)\,\delta(p+q)
\]
noting that,
\begin{align*}
\frac{\delta^{2}W[J]}{\delta J(p)\delta J(q)} & =\frac{\delta}{\delta J(p)}\,\left(\frac{1}{Z[J]}\int\mathcal{D}\phi\,e^{-S[\phi,t]+\int_{p}J_{-p}\phi_{p}}\phi(q)\right)\\
 & =\frac{1}{Z[J]}\int\mathcal{D}\phi\,e^{-S[\phi,t]+\int_{p}J_{-p}\phi_{p}}\phi(p)\phi(q)-\\
 & -\frac{\delta W[J]}{\delta J(p)}\frac{\delta W[J]}{\delta J(q)}
\end{align*}
implies that,
\[
\frac{1}{Z[J]}\int\mathcal{D}\phi\,e^{-S[\phi]-\Delta S_{\Lambda}[\phi]+\int_{p}J_{-p}\phi_{p}}\phi(p)\phi(q)=\frac{\delta^{2}W[J]}{\delta J(p)\delta J(q)}+\frac{\delta W[J]}{\delta J(p)}\frac{\delta W[J]}{\delta J(q)}\,.
\]
Thus\footnote{It is worth noting that the functional integral in (\ref{eq:dtw})
only serves to motivate the following two equations,
\begin{align*}
\partial_{t}W[J] & =\frac{1}{2}\int_{p,q}\langle\phi(p)\phi(q)\rangle\partial_{t}G_{>}^{-1}(p,q,t)\\
\langle\phi(p)\phi(q)\rangle & =\frac{\delta^{2}W[J]}{\delta J(p)\delta J(q)}+\frac{\delta W[J]}{\delta J(p)}\frac{\delta W[J]}{\delta J(q)}
\end{align*}
or the equation,
\[
\partial_{t}W[J]=\frac{1}{2}\int_{p,q}\left(\frac{\delta^{2}W[J]}{\delta J(p)\delta J(q)}+\frac{\delta W[J]}{\delta J(p)}\frac{\delta W[J]}{\delta J(q)}\right)\,\partial_{t}G_{>}^{-1}(p,q,t)
\]
these relations follow by saying that the only $u_{n}$ function that
depends on $t$ is $u_{2}(p_{1},p_{2},t)$ through $G_{>}^{-1}(p_{1},t)$,
as equation (\ref{eq:u2t}) shows and that $W[J]$ is the generating
functional of connected Green functions.},
\begin{align*}
\partial_{t}W[J] & =-\frac{1}{2}\int_{p,q}\left(\frac{\delta^{2}W[J]}{\delta J(p)\delta J(q)}+\frac{\delta W[J]}{\delta J(p)}\frac{\delta W[J]}{\delta J(q)}\right)\,\partial_{t}G_{>}^{-1}(p,q,t)\\
 & =-\frac{1}{2}\text{{Tr}}\left[\left(\frac{\delta^{2}W[J]}{\delta J\delta J}+\frac{\delta W[J]}{\delta J}\frac{\delta W[J]}{\delta J}\right)\cdot\partial_{t}G_{>}^{-1}\right]
\end{align*}
where the last line is written in terms of the trace and convolution
product of operator kernels,
\[
\text{{Tr}}[\Delta]=\int_{p}\Delta(p,p)\;,\;\Delta\cdot\Gamma(p,k)=\int_{q}\Delta(p,q)\Gamma(q,k)\,.
\]
Next, the generating functional of 1-particle irreducible vertices
is considered, it is defined by,
\[
\Gamma[\phi]=\int_{p}J_{-p}\phi_{p}-W[J]-S_{2}^{>}[\phi,t]
\]
where $J$ should be considered as a function of $\phi$ by,
\[
J=\frac{\delta\Gamma[\phi]}{\delta\phi}\,.
\]
Noting that,
\begin{align*}
\left.\partial_{t}W_{\Lambda}[J]\right|_{\phi} & =\left.\partial_{t}W_{\Lambda}[J]\right|_{J}+\int_{p}\left.\frac{\delta W[J]}{\delta J(p)}\right|_{\phi}\partial_{t}J(-p)|_{\phi}\\
 & =\left.\partial_{t}W_{\Lambda}[J]\right|_{J}+\int_{p}\phi(p)\partial_{t}J(-p)|_{\phi}
\end{align*}
leads to,
\begin{align}
\partial_{t}\Gamma[\phi] & =-\left.\partial_{t}W_{\Lambda}[J]\right|_{\phi}+\int_{p}\phi(p)\partial_{t}J(-p)|_{\phi}-\partial_{t}S_{2}^{>}[\phi,t]\nonumber \\
 & =-\left.\partial_{t}W_{\Lambda}[J]\right|_{J}-\partial_{t}S_{2}^{>}[\phi,t]\nonumber \\
 & =\frac{1}{2}\text{{Tr}}\left[\left(\frac{\delta^{2}W[J]}{\delta J\delta J}+\frac{\delta W[J]}{\delta J}\frac{\delta W[J]}{\delta J}\right)\cdot\partial_{t}G_{>}^{-1}\right]-\partial_{t}S_{2}^{>}[\phi,t]\nonumber \\
 & =\frac{1}{2}\text{{Tr}}\left[\frac{\delta^{2}W[J]}{\delta J\delta J}\cdot\partial_{t}G_{>}^{-1}\right]=\frac{1}{2}\text{{Tr}}\left[\left(\frac{\delta^{2}\Gamma_{\Lambda}[\phi]}{\delta\phi\delta\phi}+G_{>}^{-1}\right)^{-1}\cdot\partial_{t}G_{>}^{-1}\right]\nonumber \\
 & =\frac{1}{2}\text{{Tr}}\left[G_{>}\left(1+G_{>}\cdot\frac{\delta^{2}\Gamma_{\Lambda}[\phi]}{\delta\phi\delta\phi}\right)^{-1}\cdot\partial_{t}G_{>}^{-1}\cdot\right]\label{eq:dtgamma}\\
 & =-\frac{1}{2}\text{{Tr}}\left[\partial_{t}G_{>}\cdot\left(1+G_{>}\cdot\frac{\delta^{2}\Gamma_{\Lambda}[\phi]}{\delta\phi\delta\phi}\right)^{-1}\cdot G_{>}^{-1}\right]
\end{align}
where in the last equality the identity,
\[
\partial_{t}G_{>}^{-1}\cdot G_{>}=-G_{>}^{-1}\cdot\partial_{t}G_{>}
\]
was employed. Next the ERG equation for the $n$-point proper function,
\[
\Gamma(p_{1},\cdots,p_{n})=\left.\frac{\delta\Gamma[\phi]}{\delta\phi(p_{1})\cdots\delta\phi(p_{n})}\right|_{\phi=0}
\]
 is considered, (\ref{eq:dtgamma}) leads to,
\begin{equation}
\partial_{t}\Gamma(p_{1},\cdots,p_{n})=-\frac{1}{2}\text{{Tr}}\left[\partial_{t}G_{>}\cdot\frac{\delta}{\delta\phi(p_{1})\cdots\delta\phi(p_{n})}\left(1+G_{>}\cdot\frac{\delta^{2}\Gamma_{\Lambda}[\phi]}{\delta\phi\delta\phi}\right)^{-1}\cdot G_{>}^{-1}\right]_{\phi=0}\label{eq:dtg}
\end{equation}
noting that,
\[
\left(1+G_{>}\cdot\frac{\delta^{2}\Gamma_{\Lambda}[\phi]}{\delta\phi\delta\phi}\right)^{-1}=\sum_{l=0}^{\infty}\frac{1}{l!}\left(-G_{>}\cdot\frac{\delta^{2}\Gamma_{\Lambda}[\phi]}{\delta\phi\delta\phi}\right)^{.l}
\]
then,

\begin{align*}
\frac{\delta}{\delta\phi(p_{1})\cdots\delta\phi(p_{k})}\,\left(1+G_{>}\cdot\frac{\delta^{2}\Gamma_{\Lambda}[\phi]}{\delta\phi\delta\phi}\right)_{\phi=0}^{-1} & =\frac{\delta}{\delta\phi(p_{1})\cdots\delta\phi(p_{k})}\sum_{l=0}^{\infty}\frac{1}{l!}\left(-G_{>}\cdot\frac{\delta^{2}\Gamma_{\Lambda}[\phi]}{\delta\phi\delta\phi}\right)_{\phi=0}^{.l}\\
 & =\sum_{l=0}^{\infty}(-1)^{l}\sum_{m_{1}\cdots m_{l}}G_{>}\cdot\frac{\delta^{m_{1}}}{\delta\phi(p_{1})\cdots\delta\phi(p_{m_{1}})}\left(\frac{\delta^{2}\Gamma_{\Lambda}[\phi]}{\delta\phi\delta\phi}\right)\\
 & \cdots G_{>}\cdot\frac{\delta^{m_{l}}}{\delta\phi(p_{k-m_{l}})\cdots\delta\phi(p_{m_{l}})}\left(\frac{\delta^{2}\Gamma_{\Lambda}[\phi]}{\delta\phi\delta\phi}\right)
\end{align*}

\noindent where,
\[
m_{1}+m_{2}+\cdots+m_{l}=k
\]
and the summation is over all possible ways of separating the $k$
momenta into $l$ sets, the set $j$ consisting of the $m_{j}$ momenta
$k_{1}^{(j)},\cdots,k_{m_{j}}^{(j)}$. In the above equations, any
of the $m_{i}$ can be zero, thus implying that even for finite values
of $k$ the summation over $l$ has infinite terms. It is noted that
all the terms involving vanishing $m_{i}$'s implies that all the
lines connecting proper functions with more than two legs are complete
$2$ point complete propagators. This can be avoided by defining,
\[
\bar{G}_{>}(p,t)=G_{>}(p,t)\left(1+\tilde{\Gamma}_{2}(p)G_{>}(p,t)\right)^{-1}
\]
then $\bar{\ensuremath{G}}$ is the complete propagator with $\tilde{\Gamma}_{2}$
being the proper correction, and $\bar{\Gamma}_{2}$ given by,
\[
\bar{\Gamma}_{2}=\bar{G}_{>}(p,t)^{-1}
\]
is the complete proper function, which will be known when solving
the corresponding ERG equation. This amounts to restrict the summation
only over $m_{i}>0$. Replacing in (\ref{eq:dtg}) gives,
\begin{align}
\partial_{t}\bar{\Gamma}(p_{1},\cdots,p_{n}) & =-\frac{1}{2}\text{{Tr}}\left[\partial_{t}\bar{G}_{>}\cdot\sum_{l=0}^{\infty}(-1)^{l}\sum_{m_{1}\cdots m_{l}}\bar{G}_{>}\cdot\frac{\delta^{m_{1}}}{\delta\phi(p_{1})\cdots\delta\phi(p_{m_{1}})}\left(\frac{\delta^{2}\bar{\Gamma}_{\Lambda}[\phi]}{\delta\phi\delta\phi}\right)\right.\nonumber \\
 & \left.\cdots\bar{G}_{>}\cdot\frac{\delta^{m_{l}}}{\delta\phi(p_{n-m_{l}})\cdots\delta\phi(p_{m_{l}})}\left(\frac{\delta^{2}\bar{\Gamma}_{\Lambda}[\phi]}{\delta\phi\delta\phi}\right)\cdot\bar{G}_{>}^{-1}\right]_{\phi=0}\nonumber \\
 & =-\frac{1}{2}\text{{Tr}}\left[\partial_{t}\bar{G}_{>}\cdot\sum_{l=0}^{\infty}(-1)^{l}\sum_{m_{1}\cdots m_{l}}\frac{\delta^{m_{1}}}{\delta\phi(p_{1})\cdots\delta\phi(p_{m_{1}})}\left(\frac{\delta^{2}\bar{\Gamma}_{\Lambda}[\phi]}{\delta\phi\delta\phi}\right)\right.\nonumber \\
 & \left.\cdots\bar{G}_{>}\cdot\frac{\delta^{m_{l}}}{\delta\phi(p_{n-m_{l}})\cdots\delta\phi(p_{m_{l}})}\left(\frac{\delta^{2}\bar{\Gamma}_{\Lambda}[\phi]}{\delta\phi\delta\phi}\right)\right]_{\phi=0}\label{eq:wei}
\end{align}
where,
\begin{align*}
\bar{\Gamma}_{\Lambda}(p_{1}\cdots,p_{n}) & =\bar{\Gamma}_{\Lambda}(p_{1}\cdots,p_{n})\;,\forall n\neq2\\
\bar{\Gamma}_{2} & =\bar{G}_{>}(p,t)^{-1}
\end{align*}
the last equality in (\ref{eq:wei}) follows using the cyclic property
of the trace and the fact that $\bar{G}_{>}^{-1}$ commutes with $\partial_{t}\bar{G}_{>}$,
thus the $\bar{G}_{>}^{-1}$ cancels the $\bar{G}_{>}$ appearing
in the first factor. Equation (\ref{eq:wei}) is Weinberg's $1$-loop
ERG equation\footnote{Compared to Weinberg, there is a sign difference on the r.h.s. of
the ERG equation. This comes because in Weinberg's approach the derivative
respect to $\Lambda$ of $\theta(|p|-\Lambda)$ gives $-\delta(|p|-\Lambda)$.
This does not happen for the approach with dimensional regularization,
as Appendix E shows. }. The multi-index version of the equation is,
\begin{align}
\partial_{t}\bar{\Gamma}(p_{1},\cdots,p_{n}) & =-\frac{1}{2}\,\left[\partial_{t}\bar{G}_{>ij}\cdot\sum_{l=0}^{\infty}(-1)^{l}\sum_{m_{1}\cdots m_{l}}\frac{\delta^{m_{1}}}{\delta\phi(p_{1})\cdots\delta\phi(p_{m_{1}})}\left(\frac{\delta^{2}\bar{\Gamma}_{\Lambda}[\phi]}{\delta\phi\delta\phi}\right)_{jk_{1}}\right.\label{eq:wei-1}\\
 & \left.\cdots\frac{\delta^{m_{l}}}{\delta\phi(p_{k-m_{l}})\cdots\delta\phi(p_{m_{l}})}\left(\bar{G}_{>}\cdot\frac{\delta^{2}\bar{\Gamma}_{\Lambda}[\phi]}{\delta\phi\delta\phi}\right)_{k_{l-1}i}\right]_{\phi=0}
\end{align}
where the indices $i,j,\cdots$ indicate type of field, component
of field and momentum.

Alternatively, the following approach employed in Weinberg's paper
\cite{Weinberg1978} leads to the same results. Assume that all the
couplings appearing in $S[\phi,t]$ depend on $t$ in such a way that
the correlators are independent of $t$, that is, such that,
\begin{equation}
\partial_{t}W[J]=0\quad\text{{for}}\;\phi=\frac{\delta W[J]}{\delta J}\quad\text{{fixed},}\label{eq:indcorr}
\end{equation}
and that only low momentum degrees of freedom are integrated over,
this amounts to separate the $u_{2}(p_{1},p_{2},t)$ coefficient as
previously but integrating only over IR degrees of freedom, i.e.,
\[
u_{2}(p_{1},p_{2},t)=\tilde{u}_{2}(p_{1},p_{2})+G_{<}^{-1}(p_{1},t)\delta(p_{1}+p_{2})\;,\;G_{<}(p_{1},t)=\frac{\theta_{\epsilon}(\mu-|p|)}{p^{2}+m^{2}}
\]
this also implies that,
\begin{align*}
S_{2}^{<} & =\frac{1}{2}\int_{p_{1},p_{2}}G_{<}^{-1}(p_{1},t)\delta(p_{1}+p_{2})\phi(p_{1})\phi(p_{2})\\
 & =\frac{1}{2}\int_{p<\mu}G^{-1}(p)\phi(p)\phi(-p)\\
 & =\frac{1}{2}\int_{p}G^{-1}(p)\phi(p)\phi(-p)-\frac{1}{2}\int_{p>\mu}G^{-1}(p)\phi(p)\phi(-p)\\
 & =\frac{1}{2}\int_{p}G^{-1}(p)\phi(p)\phi(-p)-S_{2}^{>}
\end{align*}
and thus,
\[
\partial_{t}S_{2}^{<}=-\partial_{t}S_{2}^{>}
\]
 then in this approach,
\[
0=\partial_{t}W[J]=\langle-\partial_{t}\left(\frac{1}{2}\int_{p,q}\phi(p)G_{<}^{-1}(p,q,t)\phi(q)\right)-\partial_{t}\tilde{S}[\phi]\,\rangle
\]
where mean values are defined by,
\[
\langle A[\phi]\rangle=\frac{1}{Z[J]}\int\mathcal{D}\phi\,A[\phi]e^{-S[\phi,t]+\int_{p}J_{-p}\phi_{p}}
\]
noting that,
\begin{align*}
\left.\partial_{t}W_{\Lambda}[J]\right|_{\phi} & =\left.\partial_{t}W_{\Lambda}[J]\right|_{J}+\int_{p}\frac{\delta\partial_{t}W[J]}{\delta J(p)}\partial_{t}J(-p)|_{\phi}\\
 & =\left.\partial_{t}W_{\Lambda}[J]\right|_{J}+\int_{p}\phi(p)\partial_{t}J(-p)|_{\phi}
\end{align*}
thus (\ref{eq:indcorr}) implies that,
\begin{align*}
0=\left.\partial_{t}W_{\Lambda}[J]\right|_{\phi} & =\left.\partial_{t}W_{\Lambda}[J]\right|_{J}+\int_{p}\phi(p)\partial_{t}J(-p)|_{\phi}\\
 & =-\frac{1}{2}\int_{p,q}\left(\frac{\delta^{2}W[J]}{\delta J(p)\delta J(q)}+\frac{\delta W[J]}{\delta J(p)}\frac{\delta W[J]}{\delta J(q)}\right)\partial_{t}G_{<}^{-1}(p,q,t)-\partial_{t}\langle\tilde{S}[\phi]\rangle\\
 & =\frac{1}{2}\int_{p,q}\left(\frac{\delta^{2}W[J]}{\delta J(p)\delta J(q)}+\frac{\delta W[J]}{\delta J(p)}\frac{\delta W[J]}{\delta J(q)}\right)\partial_{t}G_{>}^{-1}(p,q,t)-\partial_{t}\langle\tilde{S}[\phi]\rangle
\end{align*}
where in the last line it was used that,
\[
\partial_{t}G_{<}^{-1}(p,q,t)=-\partial_{t}G_{>}^{-1}(p,q,t)
\]
then identifying,
\[
\partial_{t}\langle\tilde{S}[\phi]\rangle=\partial_{t}\Gamma-\partial_{t}S_{2}^{<}[\phi,t]
\]
leads to,
\begin{align*}
\partial_{t}\Gamma & =\frac{1}{2}\text{{Tr}}\left[\left(\frac{\delta^{2}W[J]}{\delta J\delta J}+\frac{\delta W[J]}{\delta J}\frac{\delta W[J]}{\delta J}\right)\cdot\partial_{t}G_{>}^{-1}\right]+\partial_{t}S_{2}^{<}[\phi,t]\\
 & =\frac{1}{2}\text{{Tr}}\left[\left(\frac{\delta^{2}W[J]}{\delta J\delta J}+\frac{\delta W[J]}{\delta J}\frac{\delta W[J]}{\delta J}\right)\cdot\partial_{t}G_{>}^{-1}\right]-\partial_{t}S_{2}^{>}[\phi,t]
\end{align*}
which is equation (\ref{eq:dtgamma}).

\section*{Appendix G. Multiplicities\label{sec:G-mul}}

In evaluating the sum on the r.h.s. of the ERG equations in Fig. \ref{fig:erg},
certain terms appear repeatedly due to the summation over indices
associated with the functional derivatives. To organize this calculation
systematically, we group terms by associating them with diagrams,
each assigned a multiplicity factor that indicates how often the term
appears in the sum. These diagrams are characterized by a specific
number of external legs, determined by the l.h.s. of the ERG equations,
and an internal structure that depends on the number of vertices.
The multiplicity factor is computed as the product of two contributions:
the number of ways to assign the external legs and the number of possible
arrangements of the internal lines within the diagram. Below, we illustrate
how to compute the multiplicity for a given diagram, and in Table
1, we summarize the results for all diagrams relevant to the equations. 

Example 1:\includegraphics[bb=0bp 1.498016cm 294.1218bp 193.5437bp,scale=0.25]{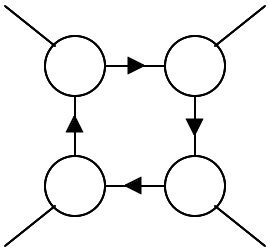} 

\begin{align*}
\partial_{t}\,\frac{\delta^{4}\Gamma}{\delta\phi_{1}\delta\phi_{2}\delta\phi_{3}\delta\phi_{4}} & =-\frac{1}{2}\,(-1)^{4}\frac{\delta^{4}}{\delta\phi_{1}\delta\phi_{2}\delta\phi_{3}\delta\phi_{4}}\\
 & \big[\,\partial_{t}G_{\bar{\psi}\psi}\,\Gamma_{\bar{\psi}\psi}\,G_{\bar{\psi}\psi}\,\Gamma_{\bar{\psi}\psi}^{\phi_{2}\phi_{3}}\,G_{\bar{\psi}\psi}\,\Gamma_{\bar{\psi}\psi}^{\phi_{3}\phi_{4}}\,G_{\bar{\psi}\psi}\,\Gamma_{\bar{\psi}\psi}^{\phi_{4}\phi_{1}}\\
 & +\,\partial_{t}G_{\psi\bar{\psi}}\,\Gamma_{\psi\bar{\psi}}^{\phi_{1}\phi_{2}}\,G_{\psi\bar{\psi}}\,\Gamma_{\psi\bar{\psi}}^{\phi_{2}\phi_{3}}\,G_{\psi\bar{\psi}}\,\Gamma_{\psi\bar{\psi}}^{\phi_{3}\phi_{4}}\,G_{\psi\bar{\psi}}\,\Gamma_{\psi\bar{\psi}}^{\phi_{4}\phi_{1}}\,\big]
\end{align*}

\noindent the factor $2$ in the multiplicity of this diagram comes
from the two terms inside the square brackets above, the factor of
$4!$ comes from the different ways of assigning one derivative respect
to the scalar external legs to each proper vertex inside the square
bracket.

\vspace{0.2cm}

\begin{center}
\includegraphics[scale=1.1]{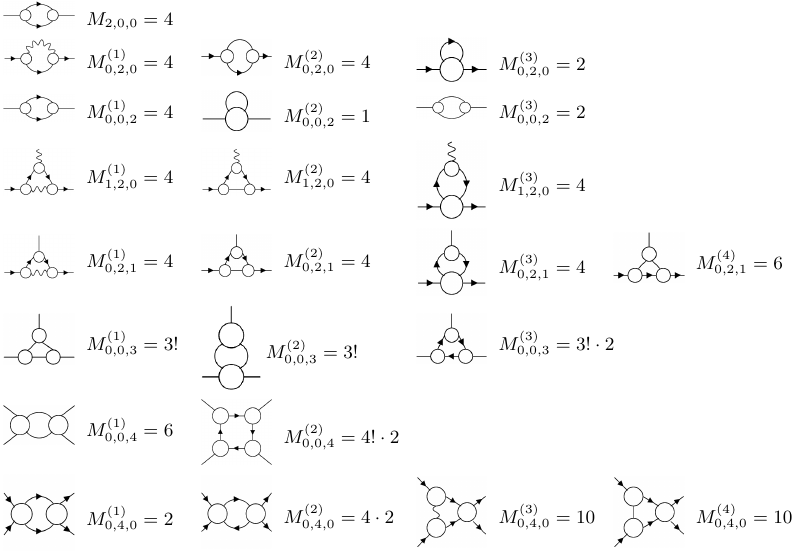}
\par\end{center}

\section*{Appendix H. The scalar potential with a cubic term.}

\subsection*{The potential}

The potential in this case is given by,
\[
V(\phi)=\mu^{2}x_{2}\phi^{2}+\mu\frac{x_{7}}{3!}\phi^{3}+\frac{x_{5}}{4!}\phi^{4}
\]
the extrema of this potential are,
\[
\phi=0,\quad\phi_{\mp}=\mu\frac{-3\text{\ensuremath{x_{7}}}\mp\sqrt{9\text{\ensuremath{x_{7}}}^{2}-24x_{2}x_{5}}}{2x_{5}}
\]
where $\phi=0$ is a maximum and for $x_{7}<0$ the absolute minimum
$\phi_{+}$ is the true vacuum and the false vacuum is given by $\phi_{-}$.
The masses associated to these minima are given by,
\begin{align*}
m_{\phi_{+}}^{2} & =\mu^{2}\,\left(-2x_{2}+\frac{x_{7}}{4x_{5}}\left(3x_{7}-\sqrt{9x_{7}^{2}-24x_{2}x_{5}}\right)\right)\,,\\
m_{\phi_{-}}^{2} & =\mu^{2}\,\left(-2x_{2}+\frac{x_{7}}{4x_{5}}\left(3x_{7}+\sqrt{9x_{7}^{2}-24x_{2}x_{5}}\right)\right)\,.
\end{align*}
The above shows that for $x_{7}<0,$ the mass associated to the absolute
minima $\phi_{+}$ is greater than the one associated to $\phi_{-}$.
It is noteworthy that this is the same that happens with the two lowest
lying positronium states, namely ortho-positronium (op) and para-positronium
(pp). As mentioned in Subsection \ref{subsec:Discussion}, the lifetimes
and masses of these states are, 
\[
\tau_{pp}=0.12\,\text{{ns}}\;,\;m_{pp}=2m_{r}-6.8\,\text{{eV}},\;\tau_{op}=140\,\text{{ns}}\;,\;m_{op}=m_{pp}+0.001\,\text{{eV}}\,.
\]
That is, the most stable state is op with a mass a little higher than
the less stable state pp. The difference of the above masses is given
by,
\[
\frac{m_{\phi_{+}}^{2}-m_{\phi_{-}}^{2}}{\mu^{2}}=-\frac{x_{7}\sqrt{9x_{7}^{2}-24x_{2}x_{5}}}{2x_{5}}
\]
which is positive for $x_{7}<0$.

\subsection*{The electron mass}

Thus the electron mass is given by $x_{1}\,\mu$ plus the Yukawa coupling
$x_{4}$ times the v.e.v. of the field $\phi$ at the true vacuum
$\phi_{+}.$ Thus,
\[
\frac{m_{e}}{\mu}=x_{1}+x_{4}\,\left(\frac{-3\text{\ensuremath{x_{7}}}+\sqrt{9\text{\ensuremath{x_{7}}}^{2}-24x_{2}x_{5}}}{2x_{5}}\right)\,,
\]
\begin{equation}
\frac{m_{e}}{\mu}=x_{1}+x_{4}\,\left(\frac{-3x_{7}+\sqrt{9x_{7}^{2}+32x_{2}\text{\ensuremath{x_{5}}}}}{8\text{\ensuremath{x_{5}}}}\right)\:.\label{eq:emass-1}
\end{equation}
As in the case with no cubic term in the potential, the value of $x_{1}$
can be changed by a shift of the field $\phi$. This shift does not
alter the masses associated with each of the minima and has no observable
consequences. This fact can be explicitly verified.

\subsection*{The positronium}

The metastable false vacuum will be associated to the op state of
the positronium which has a lifetime of $.12\,\text{{ns}}$, the true
vacuum will be associated to the para-positronium state which has
a lifetime of $142.05\,\text{{ns}},$these states have an energy difference
of $\Delta=0.001\,\text{{eV}}$. It is remarkable that the pp state
has a bigger mass than the op state. Regarding the positronium mass
squared, this is given by the coefficient of the expansion of the
potential $V(\phi)$ around the true vacuum $\phi_{+}$, this is given
by,
\[
m_{p}^{2}=\left.\frac{\partial^{2}V(\phi)}{\partial\phi^{2}}\right|_{\phi=\phi_{-}}=\mu^{2}\,\left(-2x_{2}+\frac{x_{7}}{4x_{5}}\,\left(3x_{7}-\sqrt{9x_{7}^{2}-24x_{2}x_{5}}\right)\right)
\]
this mass at the scale $\mu=m_{r}$(i.e. $t=0$) is $2m_{r}-b$ where
$b$ the binding energy of the ortho-positronium state, given by $b=6.8\,\text{{eV}},$
\[
\left(2-\frac{b}{m_{r}}\right)^{2}=4-5\times10^{-5}=\left(-2x_{2}+\frac{x_{7}}{4x_{5}}\,\left(3x_{7}-\sqrt{9x_{7}^{2}-24x_{2}x_{5}}\right)\right)(t=0)\,.
\]

\subsection*{Positronium decay}

The main contribution to positronium decay comes from the decay of
para-positronium into two photons, its mean lifetime is $\tau\simeq0.12\,\text{{ns}}$.
The first non-trivial contribution is given by the following diagram,

\begin{figure}[H]
\includegraphics[scale=0.3]{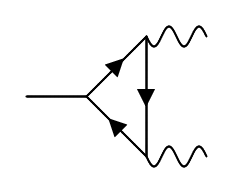}

\caption{The first contribution to positronium decay.}
\end{figure}

This diagram also appears in the fermion loop contribution to the
Higgs decay into two photons. Using the results in ref. \cite{Marciano_2012}
\cite{Slominski1667828} leads to the following expression for the
positronium decay width,
\begin{equation}
\Gamma=\frac{\alpha^{2}}{256\pi^{3}}\,\frac{x_{4}^{2}}{2}\,16\,\frac{m_{e}^{2}}{m_{p}}\,|g(\beta)|^{2}\;,\quad\beta=\frac{4m_{e}^{2}}{m_{P}^{2}}>1\label{eq:dw-1}
\end{equation}
where,
\[
g(\beta)=-2(1+(1-\beta)\,f(\beta))\;,\quad f(\beta)=\arcsin(\beta^{-1/2})\,.
\]

\subsection*{Mass difference between the false and the true vacuum}

This difference is given by,
\begin{align*}
\frac{\Delta m_{p}^{2}}{\mu^{2}} & =\frac{1}{\mu^{2}}\,\left(\left.\frac{\partial^{2}V(\phi)}{\partial\phi^{2}}\right|_{\phi=\phi_{+}}-\left.\frac{\partial^{2}V(\phi)}{\partial\phi^{2}}\right|_{\phi=\phi_{-}}\right)\\
 & =-\frac{3x_{7}\sqrt{9x_{7}^{2}-32x_{2}x_{5}}}{4x_{5}}
\end{align*}
experimentally at the scale $\mu=m_{r}$ this difference is,
\begin{equation}
\frac{\Delta m_{p}^{2}}{m_{r}^{2}}=\frac{m_{op}^{2}-m_{pp}^{2}}{m_{r}^{2}}\simeq\text{4}\times10^{-9}\label{deltam-1}
\end{equation}
where $m_{op}$ denotes the mass of ortho-positronium and $m_{pp}$
of para-positronium, the ground state. 

\subsection*{The $4$-fermion coupling}

Due to the delta function (\ref{eq:deltafipsi}), for long distances
the parameter $x_{6}$ is given in terms of $x_{2}$ and $x_{4}$
by,
\begin{equation}
x_{6}=\frac{x_{4}^{2}}{4x_{2}}\label{eq:x6-1}
\end{equation}
this relation is employed to fix the initial values of $x_{6}$ at
long distances. 

\subsection*{Initial conditions for the flow equations}

The initial conditions for the flow equations will be determined by
fixing the electron mass, its charge, the positronium mass, its decay
width and the mass difference between ortho-positronium and para-positronium
. Taking the reference scale $\mu_{0}=m_{r}$, where $m_{r}=0.51\,\text{{MeV}}$
is the electron mass at this scale, then $t=\log(\mu/\mu_{0})=0$
corresponds to $\mu=m_{r}$. The initial condition for the electron
charge is,
\begin{equation}
x_{3}(0)=-\sqrt{\frac{1}{137}}\,.\label{eq:charge-1}
\end{equation}
Regarding the electron mass, the initial condition is taken at its
rest mass $m_{r}$, which employing (\ref{eq:emass-1}) leads to,
\begin{equation}
1=x_{1}+x_{4}\,\left(\frac{-3\text{\ensuremath{x_{7}}}+\sqrt{9\text{\ensuremath{x_{7}}}^{2}-24x_{2}x_{5}}}{2x_{5}}\right)(t=0)\,.\label{eq:emass-mp-1}
\end{equation}
Regarding the positronium mass squared, this is given by the second
order coefficient in the expansion of the potential $V(\phi)$ around
the false vacuum $\phi_{+}$, this is given by,
\[
m_{p}^{2}=\left.\frac{\partial^{2}V(\phi)}{\partial\phi^{2}}\right|_{\phi=\phi_{+}}=-\mu^{2}\,\left(2x_{2}+\frac{x_{7}\left(-3x_{7}+\sqrt{9x_{7}^{2}-24x_{2}x_{5}}\right)}{4x_{5}}\right)\,.
\]
This mass, at the scale $\mu=m_{r}$(i.e. $t=0$) is $2m_{r}-b$,
where $b$ the binding energy of the positronium ground state, given
by $b=6.8\,\text{{eV}}$. This leads to:
\begin{equation}
4-3\times10^{-5}=\left(-4x_{2}-\frac{3x_{7}\left(\sqrt{9x_{7}^{2}-32x_{2}x_{5}}-3x_{7}\right)}{8x_{5}}\right)(t=0)\,.\label{eq:mp-mp-1}
\end{equation}
Regarding the positronium decay width, eq. (\ref{eq:dw-1}) will be
employed to obtain the initial value of $x_{4}$ at the scale $\mu=m_{r}$.
This leads to,
\begin{equation}
x_{4}^{2}=\Gamma\,\frac{m_{p}}{m_{e}^{2}}\,\frac{256\pi^{3}}{8\alpha^{2}|g(\beta)|^{2}}\,.\label{eq:decayp-1}
\end{equation}
Thus, there are six equations (\ref{deltam-1}), (\ref{eq:charge-1}),(\ref{eq:emass-mp-1}),(\ref{eq:mp-mp-1}),
(\ref{eq:decayp-1}) and (\ref{eq:x6-1}) for the six required initial
conditions. 

\subsection*{The flow equations and critical points}

The flow equations for the dimension-less parameters $x_{2},x_{3},x_{4},x_{5},x_{6}$
and $x_{7}$ are, 
\begin{align*}
\dot{x}_{1} & =-x_{1}+\frac{x_{1}^{3}x_{6}}{4\pi^{2}}+\frac{2x_{1}x_{3}^{2}}{\pi^{2}}-\frac{x_{1}x_{4}^{2}}{2\pi^{2}}+\frac{x_{3}^{2}}{8\pi^{2}}\,,\\
\dot{x}_{2} & =-2x_{2}+\frac{3x_{1}^{2}x_{4}^{2}}{8\pi^{2}}+\frac{x_{2}x_{5}}{32\pi^{2}}+\frac{x_{4}^{2}}{4\pi^{2}}-\frac{x_{7}^{2}}{16\pi^{2}}\,,\\
\dot{x}_{3} & =\frac{x_{3}^{3}}{12\pi^{2}}-\frac{x_{3}\text{\ensuremath{x_{4}^{2}}}}{48\pi^{2}}\,,\\
\dot{x}_{4} & =-\frac{9x_{1}^{2}x_{4}x_{6}}{8\pi^{2}}+\frac{x_{3}^{2}x_{4}}{6\pi^{2}}-\frac{x_{4}^{3}}{24\pi^{2}}\,,\\
\dot{x}_{5} & =\frac{x_{4}^{4}}{2\pi^{2}}+\frac{x_{4}^{2}x_{5}}{2\pi^{2}}-\frac{3x_{5}^{2}}{16\pi^{2}}\,,\\
\dot{x}_{6} & =2x_{6}-\frac{25\text{\ensuremath{x_{1}^{2}}}\text{\ensuremath{x_{6}^{2}}}}{16\pi^{2}}-\frac{x_{3}^{2}x_{6}}{6\pi^{2}}+\frac{5x_{4}^{2}x_{6}}{48\pi^{2}}\,,\\
\dot{x}_{7} & =x_{7}+\frac{3x_{1}x_{4}^{3}}{2\pi^{2}}+\frac{3x_{4}^{2}x_{7}}{8\pi^{2}}\,.
\end{align*}
The critical points are obtained equating to zero all the beta functions.
The only non-trivial solution is,
\[
x_{1}=x_{2}=x_{3}=x_{4}=x_{5}=x_{6}=x_{7}=0\,.
\]
The flow equations are first order ordinary non-linear differential
equations, they can be solved numerically. The results are practically
the same as for the case with $x_{7}=0$, as was expected from the
discussion in subsection \ref{subsec:Discussion}. The additional
information respect to that case is the running of the cubic coupling
$\lambda_{3}=m_{r}x_{7}$ , which is shown in the following figure,
\begin{figure}[H]
\begin{centering}
\includegraphics[scale=1.1]{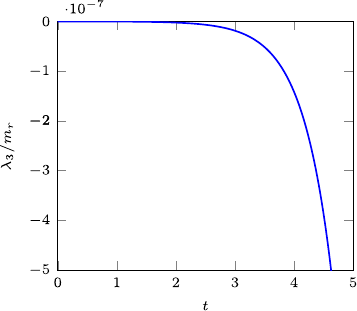}
\par\end{centering}
\caption{The cubic coupling as a function of $t=\log(\mu/m_{r})$.}
\end{figure}

\pagebreak{}

\noindent\bibliographystyle{unsrt}
\bibliography{../../Bibliography}

\end{document}